\documentclass[reprint,
 amsmath,amssymb,
 aps,superscriptaddress]{revtex4-2}
\usepackage{amsmath}
\usepackage[percent]{overpic}
\usepackage{graphicx}
\usepackage{dcolumn}
\usepackage{xcolor}
\usepackage{tikz}
\usetikzlibrary{quantikz2}
\usepackage[colorlinks=true,citecolor=blue]{hyperref}
\newcommand{\figref}[2]{\hyperlink{figtop:#1}{Fig.~\ref*{fig:#1}#2}}
\usepackage[normalem]{ulem}

\newcommand{\rev}[1]{{\color{magenta}#1}}
\renewcommand{\rev}[1]{#1}

\newcommand{\orcid}[1]{\href{https://orcid.org/#1}{\includegraphics[width=10pt]{orcid.pdf}}}
\usepackage{bm}
\usepackage{geometry}
\usepackage{braket}

\begin{document}
\title{Deterministic Universal Logical Gates for Finite-Energy GKP Qubits in Trapped Neutral Atoms}
\author{Alok Kumar}
\thanks{These authors contributed equally to this work.}
\affiliation{Department of Physics}
\author{Aaron N. Raja}
\thanks{These authors contributed equally to this work.}
\affiliation{Department of Physics}
\author{Diksha Thapliyal}
\affiliation{Department of Physics}
\author{Ishitwa Kumar Das}
\affiliation{Department of Physics}
\author{Ajay Wasan}
\thanks{Contact author: awasan@ph.iitr.ac.in}
\affiliation{Department of Physics}
\affiliation{Centre for Photonics and Quantum Communication Technology\\ Indian Institute of Technology Roorkee, Roorkee-247667, India}

\date{\today}

\begin{abstract}
 We present a universal logical gate set for finite-energy
Gottesman-Kitaev-Preskill (GKP) qubits encoded in the harmonic motional
states of trapped neutral atoms. Internal electronic states serve as an ancilla for implementing state-dependent conditional displacements in phase space, enabling arbitrary single-qubit phase gates. Transient Rydberg-excitation-mediated atomic dipole-dipole interactions enable a controlled-$Z$ gate without invoking the blockade mechanism. The gates operate under magic-trapping conditions, allowing continuous trapping throughout the gate sequence, thereby preserving the motional encoding without intermediate measurements or feedforward.
We assess the
experimental feasibility of the proposed protocols using neutral
$^{88}\mathrm{Sr}$ atoms and obtain 
average gate fidelities of $0.986$ for single-qubit phase gates
and $0.985$ for the controlled-$Z$ gate at a finite squeezing
parameter of $\Delta=0.25$ (12 dB of squeezing), increasing to
$0.997$ and $0.995$, respectively as $\Delta$ reaches $0.1$. These results provide a route toward universal, measurement-free logical control of motional GKP qubits in neutral-atom architectures.
\end{abstract}

\maketitle
\section{\label{sec:level1}Introduction}

Bosonic quantum error-correcting codes
exploit the infinite-dimensional Hilbert space of harmonic oscillators to encode logical qubits that can be continuously error corrected, providing a pathway towards fault-tolerant quantum computation~\cite{campbell2017roads,albert2018performance}. Among the various bosonic quantum error-correcting codes, the GKP code~\cite{gottesman2001encoding} has emerged as one of the most attractive options because it efficiently corrects dominant hardware errors and offers a rich set of logical operations~\cite{grimsmo2021gkp,brady2024advances}.  

Significant progress has recently been made toward the experimental realization of GKP qubits across multiple quantum computing platforms. High-fidelity GKP state preparation and control have been demonstrated in superconducting microwave cavities ~\cite{campagne-ibarcq2020qec,eickbusch2022fast,sivak2023realtime,lachance-quirion2024autonomous} and trapped ions ~\cite{fluhmann2019encoding,deneeve2022errorcorrection,matsos2024robust}, with universal logical gates demonstrated in the latter~\cite{Matsos2025}, \rev{and gate constructions for finite-energy GKP encodings have been analyzed in Ref.~\cite{PhysRevLett.133.100601}}. Trapped neutral atoms offer a complementary, well-suited platform~\cite{Saffman2016,Henriet2020quantumcomputing,bluvstein2024logical} for bosonic quantum computation, combining long-lived internal states, coherent motional dynamics, and tunable atom–atom interactions, with quantized motional degrees of freedom serving as bosonic modes.
Furthermore, coherent control of atomic motional states and their coupling to internal electronic states have been demonstrated experimentally~\cite{morinaga1999manipulation,bouchoule1999neutral,belmechri2013microwave,winkelmann2022direct}, providing essential ingredients for entangling gates. 

\rev{Optimal control techniques have been employed to generate motional GKP states of neutral atoms through modulation of optical lattice potentials and non-harmonic trapping potentials relevant to optical tweezers and lattices~\cite{kendell2024deterministic, Grochowski2025quantumcontrolof}}. Recent experiments have also demonstrated quantum tomography of the motional states of a single neutral atom in an optical tweezer, including the observation of Wigner function negativity and coherence of non-stationary states~\cite{brown2023time}. A significant step toward bosonic quantum computation with neutral atoms was recently reported in Ref.~\cite{bohnmann2025bosonic}, which proposes protocols for the preparation and protection of finite-energy GKP states encoded in the harmonic motional modes of neutral atoms, establishing the platform as a viable candidate for bosonic quantum error correction.
However, preparing and protecting logical qubits alone is insufficient for quantum computation; it also requires a universal set of logical gates. In particular, a complete gate scheme must implement both
single-qubit and two-qubit logical operations that are feasible with current experimental techniques while preserving the
motional encoding in neutral atoms. A corresponding scheme for universal logical gates, however, has yet to be established.

This work addresses these challenges by proposing a universal logical gate scheme for finite-energy motional GKP qubits encoded in trapped neutral atoms. The logical Hadamard gate is implemented exactly through free harmonic evolution. In contrast, arbitrary logical phase gates are realized using ancilla-motional entanglement and controlled displacements in phase space to generate the desired logical phase. Rydberg interactions are further exploited to implement a logical two-qubit controlled-Z gate. Finally, we investigate a representative implementation using $^{88}\mathrm{Sr}$ and assess the feasibility of the proposed scheme under realistic experimental conditions.

The structure of this paper is as follows. In Sec.~\ref{sec:level2}, we briefly review finite-energy GKP states and the logical operations relevant to this work. Section~\ref{sec:level5} outlines the key experimental ingredients, including magic trapping and conditional displacement operations. Section~\ref{sec:level8} presents the implementation of the logical Hadamard and arbitrary single-qubit phase gates, while Sec. ~\ref{sec:level12} introduces the Rydberg-mediated protocol for the logical controlled-$Z$ gate. In Sec.~\ref{sec:level13}, we discuss a representative implementation using neutral $^{88}$Sr atoms and identify experimentally realistic operating parameters. In Sec.~\ref{sec:level14}, we present the durations and fidelity analysis of the proposed single- and two-qubit gates. Finally, we conclude in Sec.~\ref{sec:level15} with an outlook for future work. Additional technical
details are provided in the appendixes.

\section{\label{sec:level2}GKP CODE}

In neutral-atom platforms, heating and motional noise can deteriorate quantum information encoded in the motion of trapped atoms. An effective framework for mitigating such errors is provided by the GKP code~\cite{gottesman2001encoding,brady2024advances}. Moreover, these qubits can be combined with the surface code in the surface-GKP architecture, providing a scalable route toward fault-tolerant quantum computation~\cite{noh2022surfacegkp}. In
this section, we review the basic properties of ideal and finite-energy \rev{square-lattice} GKP codes.
\begin{figure*}[t]
    \hypertarget{figtop:1}{}
    \begin{overpic}[width=0.48\textwidth]{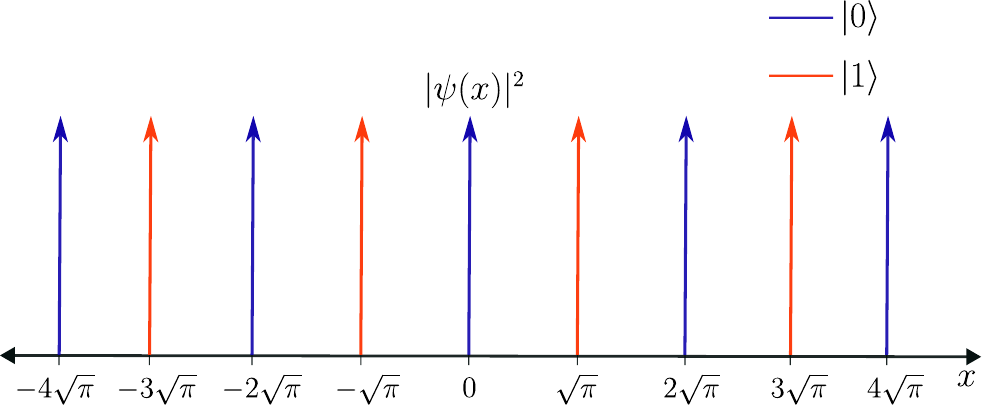}
    \put(2,40){{(a)}}
    \end{overpic}
    \hfill
    \begin{overpic}[width=0.50\textwidth]{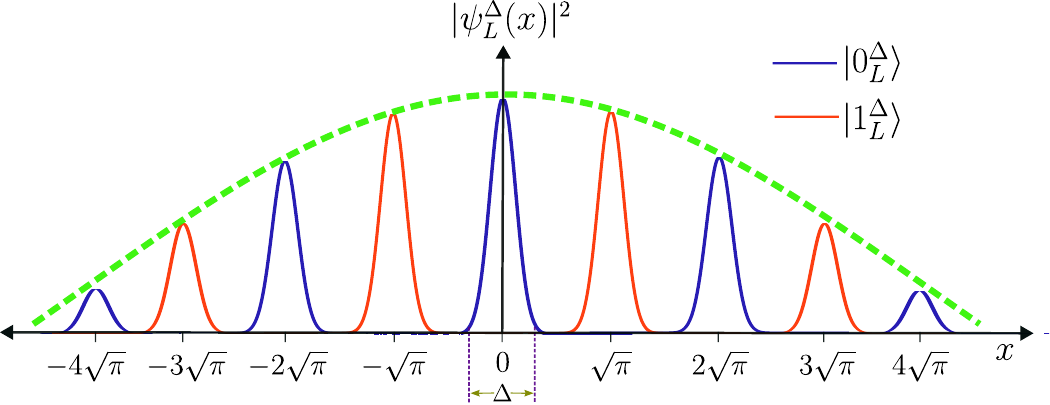}
    \put(4.8,39){{(b)}}
    \end{overpic}
    \caption{Illustration of (a) ideal and (b) finite-energy \rev{square-lattice GKP logical states} in the position quadrature. The ideal states consist of infinitely sharp peaks, whereas the finite-energy states exhibit Gaussian-broadened peaks modulated by an overall Gaussian envelope with its width determined by the squeezing parameter $\Delta$.}
\label{fig:1}
\end{figure*}
\subsection{\label{sec:level3}Ideal GKP}
In the ideal GKP code, the logical basis states are encoded as infinite combs of equally spaced eigenstates of the position and momentum quadratures \rev{as shown in \figref{1}{(a)}}. The dimensionless quadrature operators are defined as $\hat{q}=\dfrac{\hat{a}^\dagger+\hat{a}}{\sqrt{2}}$ and $\hat{p}=i\dfrac{\hat{a}^\dagger-\hat{a}}{\sqrt{2}}$ which satisfy the commutation relation [$\hat{q},\hat{p}$] = $i$. The logical basis states are defined by:
\begin{equation}
|0_L\rangle = \sum_{k=-\infty}^{\infty} |q = 2k\sqrt{\pi}\rangle,
\end{equation}
\begin{equation}
\begin{aligned}
|1_L\rangle
&= \sum_{k=-\infty}^{\infty} |q=(2k+1)\sqrt{\pi}\rangle.
\end{aligned}
\end{equation}
Logical gate operations are naturally expressed in terms of displacement and phase-space rotation operators. The phase-space displacement operation is defined as $
\hat{D}(\alpha)=\exp[(\alpha \hat{a}^\dagger-\alpha^\ast \hat{a})/\sqrt{2}]
$. Real and imaginary values of $\alpha$ correspond to displacements along
the $\hat{q}$ and $\hat{p}$ quadratures, respectively.

Free harmonic evolution for a duration $t$ generates a phase-space rotation described by $\hat{R}(t)= \exp(-i\omega t\hat{a}^\dagger \hat{a})$, where  $\omega$ is the trap frequency. Within the GKP code space, the logical Pauli operators and the Hadamard gate are implemented as
\begin{equation}
\hat{{X}}_L=\hat{D}(\sqrt{\pi}),\hspace{0.5cm}\hat{{Z}}_L=\hat{D}(i\sqrt{\pi}),
\end{equation}
\begin{equation}
\hat{{Y}}_L=i\hat{{X}}_L\hat{{Z}}_L,\hspace{0.5cm}\hat{{H}}_L=\hat{R}({\pi}/2\omega).
\end{equation}
\rev{The stabilizer group $\mathcal{S}$ of the ideal square GKP code is generated by $\hat{S}_Z=\hat{D}(i2\sqrt{\pi})$ and $\hat{S}_X=\hat{D}(2\sqrt{\pi})$:
    \begin{equation}
        \mathcal{\hat{S}}
        = \left\{
            \hat{S}_Z^k\hat{S}_X^l
            \,\middle|\, k,l\in\mathbb{Z}
          \right\}.
    \end{equation}} 
The logical state $|\psi_L\rangle$ is its $+1$ eigenstate, satisfying
\begin{equation}
\mathcal{\hat{S}}|\psi_L\rangle=|\psi_L\rangle.
\end{equation}
Since the ideal GKP codewords possess infinite energy and are therefore unphysical, practical implementations employ finite-energy approximations, which are discussed in the following subsection.

\subsection{\label{sec:level4}Finite GKP}

In physical implementations, the ideal GKP codewords must be replaced by finite-energy approximations. A convenient construction, \rev{illustrated in \figref{1}{(b)}}, is obtained by exponentially suppressing the occupation of high Fock states, yielding
\begin{equation}
|\psi_L^\Delta\rangle
=
2\sqrt{\pi}\,\Delta \,
e^{-\Delta^2 \hat{a}^\dagger \hat{a}}
|\psi_L\rangle,
\end{equation}
where $|\psi_L^\Delta\rangle$ and $|\psi_L\rangle$ denote the finite-energy  and ideal GKP states \rev{of the square lattice}, respectively, and $\Delta$ is the effective squeezing parameter. The ideal GKP state is obtained in the limit $\Delta \to 0$. For small values of $\Delta$, the finite-energy logical basis states are written as
\begin{equation}
\begin{aligned}
|0_L^\Delta\rangle \propto
\sum_{k=-\infty}^{\infty}
e^{-2\pi \Delta^2 k^2}
\hat{D}\!\left(2k\sqrt{\pi}\right)
\hat{S}(-\ln \Delta)\,|0\rangle\\
+ \mathcal{O}(\Delta^4),
\end{aligned}
\end{equation}

\begin{equation}
\begin{aligned}
|1_L^\Delta\rangle
&\propto
\sum_{k=-\infty}^{\infty}
e^{-\frac{\pi\Delta^2(2k+1)^2}{2}}
\hat{D}\!\left((2k+1)\sqrt{\pi}\right)
\\
&\qquad\times
\hat{S}(-\ln\Delta)\,|0\rangle
+\mathcal{O}(\Delta^4),
\end{aligned}
\end{equation}
where
\begin{equation}
\hat{S}(z)
=
\exp\!\left[
\frac{1}{2}
\left(
z^* \hat{a}^2
-
z \hat{a}^{\dagger 2}
\right)
\right]
\end{equation}
denotes the single-mode squeezing operator.
The squeezed vacuum state $\hat{S}(-\ln \Delta)|0\rangle$
has effective squeezing parameters $\Delta_Z=\Delta$
and $\Delta_X=1/\Delta$, where $\Delta_Z$ and $\Delta_X$ denote the effective squeezing associated with the logical $X$ and $Z$ stabilizer directions, respectively. The preparation protocol proposed in Ref.~\cite{bohnmann2025bosonic} enables the generation of finite-energy GKP states in a neutral-atom optical lattice, with an experimentally achievable value of
$\Delta \simeq 0.25$. 
For finite-energy GKP states, the logical Pauli operators are generalized to~\cite{royer2020stabilization}:
\begin{equation}
\begin{aligned}
\hat{\mathcal{X}}_{L}^{\Delta}
=
e^{-\Delta^{2}\hat{a}^{\dagger}\hat{a}}
\,\hat{D}(\sqrt{\pi})\,
e^{\Delta^{2}\hat{a}^{\dagger}\hat{a}},\\
\hat{\mathcal{Z}}_{L}^{\Delta}
=
e^{-\Delta^{2}\hat{a}^{\dagger}\hat{a}}
\,\hat{D}(i\sqrt{\pi})\,
e^{\Delta^{2}\hat{a}^{\dagger}\hat{a}}.
\end{aligned}
\end{equation}
These logical operators provide the basis for the logical gate constructions developed in the following sections.

\section{\label{sec:level5}Experimental Requirements and Implementation}
The implementation of the logical gates requires two key experimental ingredients: (i) magic trapping of both the ancilla and Rydberg states and (ii) ancilla state-dependent phase-space displacement operations.

The ancilla states must experience identical trapping potentials to prevent state-dependent motional evolution arising from differential AC Stark shifts~\cite{grimm2000optical}, generating unwanted ancilla-motional entanglement. This condition is realized by operating the optical lattice at a magic wavelength, where the differential Stark shift vanishes, and the motional Hamiltonian becomes identical for both ancilla states.
Conventional neutral-atom Rydberg gates typically switch off the trapping field during Rydberg excitation as the negative polarizability of the highly excited state prevents simultaneous trapping of the ancilla and Rydberg states in a red-detuned optical trap~\cite{zhang2011magic}. Such an approach is incompatible with motional GKP qubits, as removing the trap destroys the harmonic confinement required to preserve the encoded oscillator state. Therefore, the optical trap must remain continuously on, requiring both the ancilla and Rydberg states to experience identical trapping potentials. This condition can be realized in divalent atoms by exploiting the Rydberg-landscaping effect to match the Rydberg-state polarizability to that of the ancilla states through an appropriate choice of the lattice wavelength and Rydberg level~\cite{topcu2014divalent}.  

We now turn to the second key ingredient of the proposed implementation: state-dependent displacement operations.
The conditional displacement operation couples the internal ancilla state to the motional degree of freedom by producing an ancilla-state-dependent displacement of the motional GKP state in phase space. We adopt the conditional-displacement protocol introduced in Ref.~\cite{bohnmann2025bosonic} in which a state-dependent force is generated using a tune-out laser to realize the required displacement. \rev{The displacement is conditioned on the ancilla being in the $|1\rangle$ state, while no displacement is applied when the ancilla is in the $|0\rangle$ ancilla state i.e, giving the conditional evolution 
\[
|0\rangle\langle0|\otimes\hat I+
|1\rangle\langle1|\otimes\hat U(t).
\]
}The displacement amplitude $\alpha_d$ is determined by the laser parameters, with its dependence and the corresponding unitary evolution  $\hat U(t)$ presented in Appendix~\ref{sec:level18}.

\begin{figure*}[t]
\hypertarget{figtop:2}{}
\raggedright
\includegraphics[width=\linewidth,height=5.5cm]{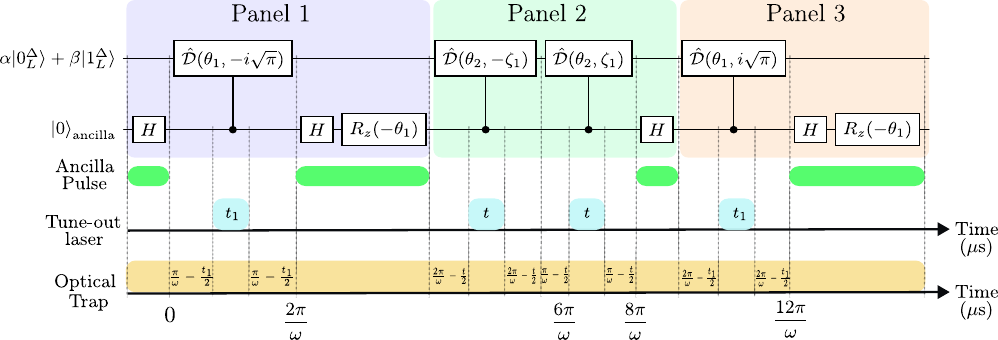}
\caption{\rev{Gate sequence for implementing the logical phase gate. In \textbf{Panel~1}, ancilla--motion entanglement is generated by applying the conditional displacement $\hat{\mathcal{D}}(\theta_1, -i\sqrt{\pi})$, followed by single-qubit rotations of the ancilla. In \textbf{Panel~2}, successive conditional displacements $\hat{\mathcal{D}}(\theta_2, -\zeta_1)$ and $\hat{\mathcal{D}}(\theta_2, \zeta_1)$ trace a closed trajectory in phase space, resulting in the accumulation of the desired logical phase. Finally, in \textbf{Panel~3}, the inverse displacement $\hat{\mathcal{D}}(\theta_1, i\sqrt{\pi})$, together with ancilla rotations, disentangles the ancilla from the motional mode. The upper and middle lines of the quantum circuit represent operations on the motional mode and the ancilla, respectively. In the pulse sequence shown below the quantum circuit, the green segments represent instantaneous ancilla pulses, the blue segments indicate tune-out laser pulses with their corresponding durations, and the brown region represents the continuously applied optical trap. The horizontal axis denotes time in $\mu$s.}}
\label{fig:2}
\end{figure*}
\section{\label{sec:level8}Gate Mechanisms for Single-Qubit Hadamard and Phase Gates}
In this section, we describe the implementation of logical single-qubit gates on finite-energy GKP qubits encoded in the motional states of trapped atoms. In particular, we consider the logical Hadamard gate and arbitrary logical phase gates.
The finite-energy GKP basis states are $|0_L^\Delta\rangle$ and $|1_L^\Delta\rangle$, while the internal electronic states $|{}^3P_0\rangle$ and $|{}^3P_2\rangle$, \rev{ which form a fine-structure qubit~\cite{PhysRevLett.132.150605},} serve as the ancilla states $|0\rangle$ and $|1\rangle$, respectively.
 Following the motional state-preparation protocol of Ref.~\cite{bohnmann2025bosonic}, the ancilla is reset to the $|0\rangle$ state, thereby ensuring that every gate sequence begins with the ancilla initialized in the $|{}^3P_0\rangle$ state~\cite{deneeve2022errorcorrection}. Throughout this work, we assume that operations can be applied selectively to the ancilla states without disturbing the motional state. \rev{Ancilla rotations can be performed on a timescale of approximately $\text{100 ns}$~\cite{unnikrishnan2024coherent}, which is much shorter than the motional oscillation period in our case. Therefore, motional evolution during the ancilla rotations can be neglected.}

As described in Sec.~\ref{sec:level3}, the Hadamard gate corresponds to a phase-space rotation, and is therefore implemented by allowing the harmonic oscillator to evolve freely for a duration of $\pi / 2\omega$. Since the free evolution operator commutes with the finite-energy envelope, the envelope is preserved under the evolution. As a result, the finite-energy GKP state undergoes the same $\pi/2$ phase-space rotation as the ideal GKP state, implementing the logical Hadamard gate exactly without introducing any additional logical distortion (see Appendix~\ref{sec:level17}).
\begin{figure*}[t]
\hypertarget{figtop:3}{}
\centering
\includegraphics[width=\linewidth]{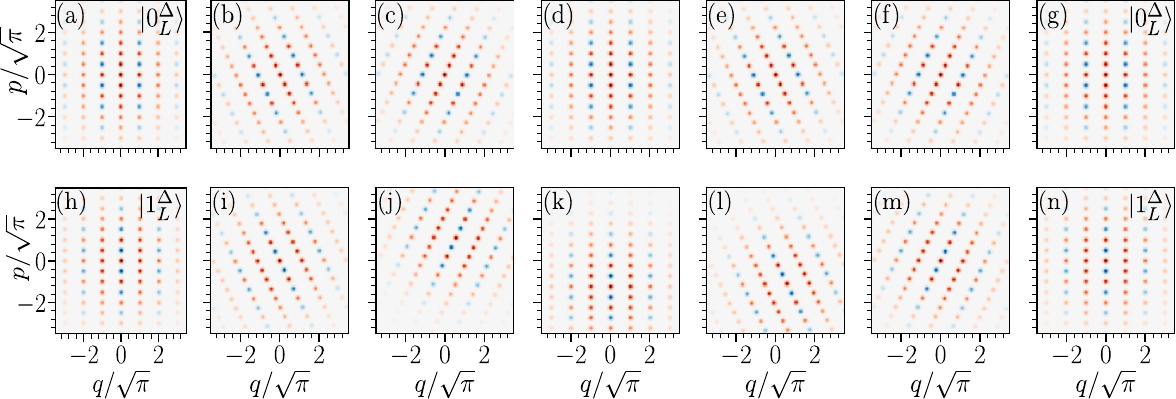}
\caption{\rev{Phase-space evolution of the logical states during the displacement-cancellation sequence} (Panel~2 of Fig.~2). The motional Wigner functions of $|0_L^\Delta\rangle|0\rangle$ (upper row, subplots~(a)--(g)) and $|1_L^\Delta\rangle|1\rangle$ (lower row, subplots~(h)--(n)) are shown for $\Delta = 0.2$ and a target phase of $\phi = \pi/4$. Red and blue indicate positive and negative values of the Wigner function, respectively. Subplots~(a) and (h) depict the initial states. Both components initially undergo a harmonic rotation $\hat{R}(\pi/\omega - t/2)$ [subplots~(b) and (i)], followed by ancilla-dependent operations: an additional rotation $\hat{R}(t)$ for the upper component [subplot~(c)] and a conditional displacement $\hat{U}(t)$ for the lower component [subplot~(j)]. A second harmonic rotation $\hat{R}(\pi/\omega - t/2)$ is then applied to both components [subplots~(d) and (k)], returning the upper component to $|0_L^\Delta\rangle|0\rangle$ while transforming the lower component into $\hat{\mathcal{D}}(\theta_2, \zeta_1)|1_L^\Delta\rangle|1\rangle$. The subsequent displacement-cancellation operations [subplots~(e)--(g) and (l)--(n)] restore both components to their respective initial phase-space configurations, with the lower component acquiring the desired relative phase $\phi$.}
\label{fig:3}
\end{figure*}
Unlike the logical Hadamard gate, which follows directly from free harmonic evolution, the logical phase gate, defined in the logical basis ($|0_L^\Delta\rangle,$ $|1_L^\Delta\rangle$) as 
$
U_{\phi}^{(L)}=
\begin{pmatrix}
1 & 0\\
0 & e^{i\phi}
\end{pmatrix}
$, 
requires ancilla-mediated composite displacement operations, \rev{conditioned on the ancilla being in the $|1\rangle$ state. These operations are defined as}
\begin{align}
\hat{\mathcal D}(\theta,-\zeta)
&\equiv
\hat R\!\left(\frac{\pi}{\omega}-\frac{ t}{2}\right)
\hat U(t)
\hat R\!\left(\frac{\pi}{\omega}-\frac{t}{2}\right)
\nonumber\\
&=
\hat D(-\zeta)
e^{i\theta},
\label{eq:10}
\end{align}
\begin{align}
\hat{\mathcal D}(\theta,\zeta)
&\equiv
\hat R\!\left(\frac{2\pi}{\omega}-\frac{ t}{2}\right)
\hat U(t)
\hat R\!\left(\frac{2\pi}{\omega}-\frac{t}{2}\right)
\nonumber\\
&=
\hat D(\zeta)
e^{i\theta},
\label{eq:11}
\end{align}
where
$\theta=\frac{\alpha_d^2}{2}\left[\omega t-\sin(\omega t)\right]$ and $\zeta\equiv 2i\alpha_d \sin\left(\frac{\omega t}{2}\right)$, both depend on t. As seen from Eqs.~\ref{eq:10} and~\ref{eq:11}, both composite operations contain the same interaction \rev{
\[
\hat{U}(t)
=
\hat{D}\!\left[\alpha_d\left(1-e^{-i\omega t}\right)\right]
e^{-i\omega \hat{a}^{\dagger}\hat{a}t}
e^{\frac{i\alpha_d^{2}}{2}
(\omega t-\sin\omega t)}\]} of duration $t$, while the surrounding harmonic rotations determine the total operation time. The derivation of these operators from $\hat U(t)$ is presented in Appendix~\ref{sec:level18}. The total durations of $\hat{\mathcal D}(\theta,-\zeta)$ and $\hat{\mathcal D}(\theta,\zeta)$ are $2\pi/\omega$ and $4\pi/\omega$, respectively.

\rev{The protocol consists of three stages as summarised in the form of quantum circuit shown in} \figref{2}{}. The first stage (Panel 1) entangles the motional GKP qubit with the internal ancilla, the second stage (Panel 2) generates the desired logical phase, and the final stage (Panel 3) disentangles the ancilla, leaving the phase entirely encoded in the motional qubit.

\subsection{\label{sec:level9}Entangling the Motional and Ancilla States}
The first stage of the protocol entangles the motional GKP qubit with the internal ancilla by associating each logical basis state with a distinct ancilla state. The corresponding gate sequence is illustrated in Panel 1 of \figref{2}{}. 

We begin with an arbitrary logical motional state $(\alpha|0_L^\Delta\rangle
+\beta|1_L^\Delta\rangle)$ while the ancilla is initialized in the $|0\rangle$ state. An ancilla-only Hadamard gate ($H_{\mathrm{anc}}$) prepares the superposition $\frac{1}{\sqrt{2}}(|0\rangle+|1\rangle)$. The composite displacement operator $\hat{\mathcal D}(\theta_1,-i\sqrt{\pi})$, conditioned on the ancilla being in the $|{}^3P_2\rangle$ state, is then applied. Here $\alpha_d$ is taken to be $2.37$ as calculated in Appendix~\ref{sec:level18}.
This displacement operator is realized by applying the tune-out laser for a duration of \rev{$t = 3.04~\mu\mathrm{s}$}, preceded and followed by an idle harmonic evolution, resulting in an accumulated phase \rev{$\theta_1=0.065\pi$} as follows from Eq.~\ref{eq:10}.
Finally, a second $H_{\mathrm{anc}}$ followed by an ancilla rotation $R_z(-\theta_1)$ compensates the known phase accumulated during the conditional displacement, yielding the entangled state
$\alpha|0_L^\Delta\rangle|0\rangle+\beta|1_L^\Delta\rangle|1\rangle$, which serves as the input to the phase-generation stage.
\subsection{\label{sec:level10}Ancilla-Conditioned Phase Generation}
The logical phase is generated by a sequence of ancilla-conditioned composite displacement operations that drive the motional state along a closed trajectory in phase space, as illustrated in Panel 2 of \figref{2}{}.  The closed trajectory produces a finite phase while returning the motional state to its initial configuration in phase space, with no net displacement. To realize this phase, we employ the  conditional displacement operators defined in Eqs.~\eqref{eq:10} and~\eqref{eq:11}.  Starting from the entangled state
$
\alpha|0_L^\Delta\rangle|0\rangle+\beta|1_L^\Delta\rangle|1\rangle,
$
$\hat{\mathcal{D}}(\theta,-\zeta_1)$ is applied conditioned on the ancilla state
$|1\rangle$. As
\[
\hat{R}\left(\frac{2\pi}{\omega}\right)|0_L^\Delta\rangle
=
|0_L^\Delta\rangle,
\]
the $|0_L^\Delta\rangle|0\rangle$ component returns to itself after
the sequence, whereas the $|1_L^\Delta\rangle|1\rangle$ component undergoes the momentum
displacement shown in \figref{3}{}, and acquires a phase $\theta$ .

The subsequent application of $\hat{D}(\theta,\zeta_1)$, conditioned on $|1\rangle$, again leaves the $|0_L^\Delta\rangle|0\rangle$ component unchanged, while the $|1_L^\Delta\rangle|1\rangle$ component undergoes a displacement opposite to that of the first operation, returning it to its initial point in phase space, as shown in \figref{3}{}. The phase accumulated during this operation adds to that from the first operation, resulting in a total phase $2\theta$ relative to the $|0_L^\Delta\rangle|0\rangle$ component.
The resulting state is
therefore
$
\alpha|0_L^\Delta\rangle|0\rangle
+
e^{i2\theta}\beta|1_L^\Delta\rangle|1\rangle.
$

\begin{figure}[t]
    \hypertarget{figtop:4}{}
    \centering
    \includegraphics[width=0.48\textwidth]{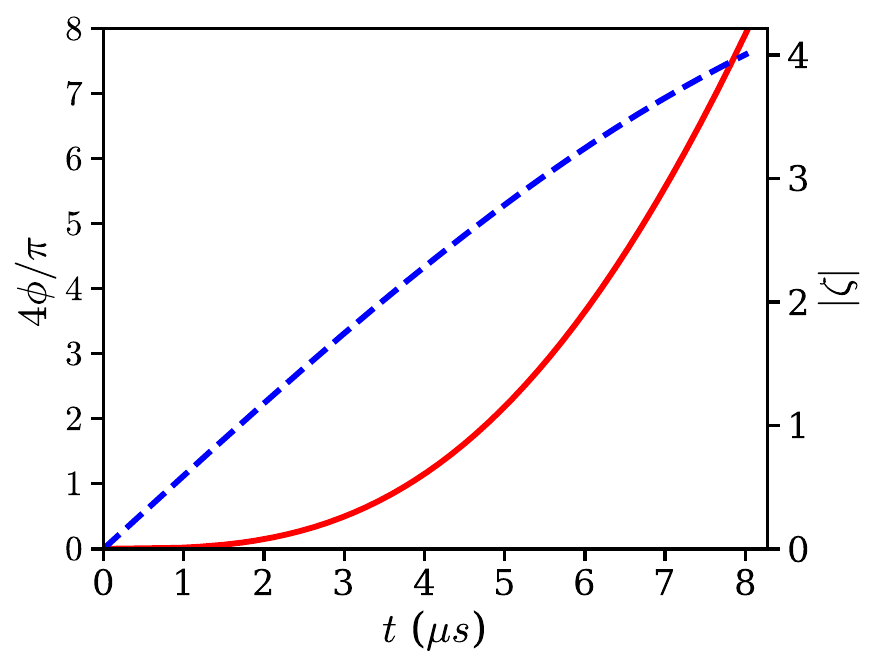}
     
    \caption{\rev{Target logical phase $\phi$ (in radians, solid curve) as a function of the duration $t$ (in $\mu$s) of the operator $\hat{U}(t)$. The larger target logical phases require longer tune-out laser pulse. The dashed curve  shows values of $|\zeta|$ from Eq.~\ref{eq:10}, which correspond to the displacement required along the $\hat{p}$ quadrature to impart each target phase.}}
    \label{fig:4}
\end{figure}

For a general logical phase gate, the duration $t$ of each application of the conditional evolution operator $\hat U(t)$ is chosen such that the total accumulated phase satisfies
$
2\theta=\phi,
$
where $\phi$ is the desired logical phase. Substituting 
$
\theta=\frac{\alpha_d^2}{2}\left[\omega t-\sin(\omega t)\right],
$ into $
2\theta=\phi,
$ gives
\[
\alpha_d^2\bigl(\omega t-\sin(\omega t)\bigr)
=
\phi
.\]
For $\alpha_d=2.37$, the accumulated phase $\phi$ as a function of the duration $t$ of the conditional evolution operator $\hat U(t)$ is shown in \figref{4}{}.

\subsection{\label{sec:level11}Disentangling the Motional and Ancilla States}
The final stage removes the ancilla-motional entanglement while preserving the logical phase accumulated during the previous step. As shown in Panel 3 of \figref{2}{}, this is accomplished by reversing the initial entangling sequence.

$H_{\mathrm{anc}}$ is first applied, followed by a  conditional displacement operation $\hat{\mathcal D}(\theta_1,i\sqrt{\pi})$ which exactly reverses the entangling operation introduced in Sec.~\ref{sec:level9}. Finally, $H_{\mathrm{anc}}$ followed by an $R_z(-\theta_1)$ rotation restores the ancilla to its initial state, yielding $(\alpha|0_L^\Delta\rangle
+
e^{i\phi}
\beta|1_L^\Delta\rangle)\otimes|0\rangle$. The ancilla is therefore completely disentangled from the motional mode, while the desired logical phase remains encoded in the finite-energy GKP qubit. The protocol therefore realizes a deterministic arbitrary logical phase gate without requiring intermediate measurement or feedforward.

\begin{figure*}[t]
    \hypertarget{figtop:5}{}
    \centering
    \begin{overpic}[width=0.58\textwidth]{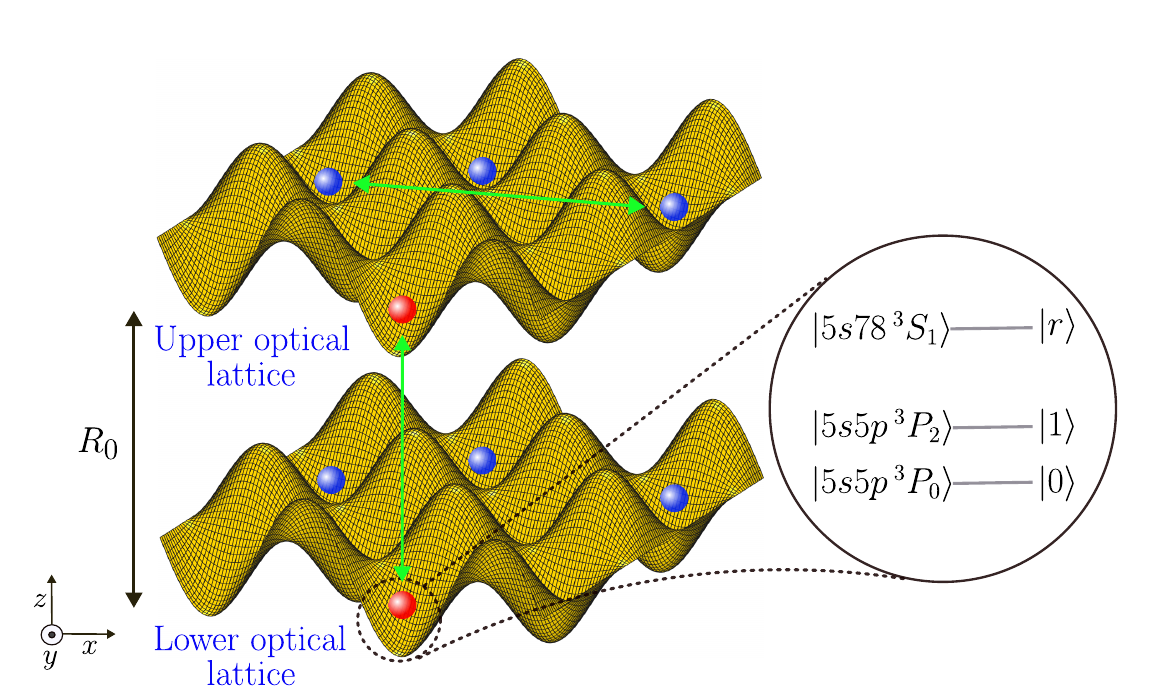}
    \put(2,50){{(a)}}
    \put(101,50){{(b)}}
    \put(128,50){{(c)}}

    \end{overpic}
     \begin{overpic}[width=0.40\textwidth]{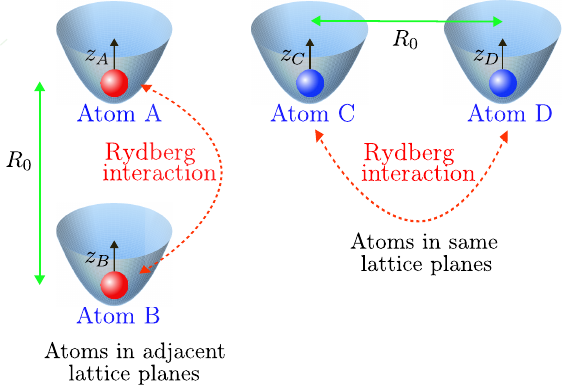}

    \end{overpic}
    \caption{\rev{Schematic of the proposed neutral-atom architecture for implementing logical gates on motional GKP qubits. (a)~Two parallel optical lattice planes separated by a distance $R_0$, with each plane sparsely populated by $^{88}\mathrm{Sr}$ atoms such that the in plane interatomic separation is at least $R_0$. The logical qubit is encoded in the axial motional mode of each atom, while the ancilla is encoded in the internal states $|0\rangle$ and $|1\rangle$, with the Rydberg state $|r\rangle$ mediating interatomic interactions. (b), (c)~Two configurations for implementing Rydberg-mediated
interactions are shown. In (b), the atoms are in adjacent lattice
planes, vertically aligned at the same $(x,y)$ position and separated
by $R_0$ along the $z$-direction. In (c), the atoms are in the same
lattice plane and separated by an in-plane distance $R_0$. In both
configurations, simultaneous Rydberg excitation of the two atoms gives
rise to an interatomic van der Waals interaction. }}
    \label{fig:5}
\end{figure*}
\section{\label{sec:level12}Gate Mechanism for\\ Controlled-$Z$ Gate}
Having established the implementation of arbitrary single-qubit logical
phase gates, an entangling logical controlled-$Z$ (CZ) gate in the logical basis $(|0_L^\Delta 0_L^\Delta\rangle,
|0_L^\Delta 1_L^\Delta\rangle,
|1_L^\Delta 0_L^\Delta\rangle,
|1_L^\Delta 1_L^\Delta\rangle)$ is defined as 
\begin{equation}
U_{\mathrm{CZ}}=\operatorname{diag}(1,1,1,-1).
\end{equation}
Similar to the logical phase gate discussed in Sec.~\ref{sec:level8}, the protocol for the CZ gate consists of three stages: ancilla-motional entanglement, phase generation using  Rydberg interaction, and ancilla-motional disentanglement.

When both atoms are  excited simultaneously to the transient
Rydberg state, forming the doubly excited state $|rr\rangle$, they interact through the van der Waals potential  $V(R)=\frac{C_6}{R^6},
$ where $R$ denotes the interatomic separation and $C_6$ is the van der Waals coefficient~\cite{saffman2010rydberg}. \rev{Rydberg-mediated gates in $^{88}$Sr involve several considerations, including Rydberg-state decay, photon recoil, and the form of the interaction potential~\cite{PhysRevResearch.4.033019}}. The present gate protocol
operates in a regime of weak Rydberg--Rydberg  interaction in which
double excitation to $|rr\rangle$ is allowed, and the resulting
interaction is deliberately exploited to generate the conditional
phase.
Throughout the gate operation, the trapping potential remains active, with the Rydberg excitation performed under the magic-trapping conditions described in Sec.~\ref{sec:level5}, thereby preserving the motional encoding during the interaction. \rev{We consider two different configurations of the interatomic separation, each of which independently enables the implementation of the CZ gate distinguished by the orientation of the equilibrium separation relative to the encoded axial motional direction. In the first configuration, atoms in adjacent lattice layers are separated along the $z$-axis, so that the equilibrium interatomic separation is aligned with the encoded motional direction, as shown in \figref{5}{(b)}. In the second configuration, atoms in the same lattice plane are separated transversely, perpendicular to the encoded axial motional direction as shown in \figref{5}{(c)}.} 

For the first configuration, the motional dynamics of the two atoms in the doubly excited state $|rr\rangle$ are governed by
\begin{equation}
H_{rr} = \sum_{i=1}^{2} \left( \frac{p_i^2}{2m} + \frac{1}{2}m\omega^2 z_i^2 \right) + \frac{C_6}{|R_0 + z_1 - z_2|^6},
\end{equation}
where $R_0$ is the equilibrium interatomic separation and $z = z_1 - z_2$ is the relative axial displacement. Introducing center-of-mass and relative coordinates and their conjugate momenta, the Hamiltonian separates into independent center-of-mass and relative-motion contributions, as discussed in Appendix~\ref{sec:level19}. Since the Rydberg interaction depends only on the relative coordinate, the center-of-mass mode remains an independent harmonic oscillator, decoupled from the interaction. We therefore focus on the effective Hamiltonian of the relative mode,
\begin{equation}
\begin{aligned}
H_{\mathrm{eff}} = &\frac{p^2}{m} + \frac{1}{4} m \omega^2 z^2 +
\frac{C_6}{R_0^6}
- \frac{6C_6}{R_0^7} z\\&
+ \frac{21C_6}{R_0^8} z^2 
+ \mathcal{O}(z^3). 
\end{aligned}
\end{equation}
For the gate dynamics, we retain the harmonic term and the constant and linear contributions from the Rydberg interaction. \rev{The constant term provides the dominant contribution to the accumulated phase, while the linear term generates an unwanted state-dependent motional displacement that must be cancelled to ensure the desired gate operation}. The quadratic term introduces motional distortion and is included separately in the fidelity analysis. The corresponding time-evolution operator is
\begin{equation}
\begin{split}
\hat V(t) = {}& \hat D\!\left[\beta_d\!\left(1-e^{-i\omega t}\right)\right]
e^{-i\omega t\hat a^\dagger\hat a} \\
&\times \exp\!\left\{i\frac{\beta_d^{2}}{2}\bigl(\omega t-\sin(\omega t)\bigr)
-i\frac{C_6 t}{R_0^{6}\hbar}\right\},
\end{split}
\end{equation}
where $
\beta_d=
\frac{6C_6}{R_0^7\,\sqrt{\hbar m\omega^3/2}}$ and the derivation of $\hat V(t)$ from $H_\mathrm{eff}$ is presented in Appendix~\ref{sec:level19}.

\rev{For the second configuration, in which the equilibrium interatomic separation lies in the transverse plane, the Rydberg interaction depends on the relative axial displacement as \(V(z)=C_6/[R_0^2+z^2]^3\).  Following the same procedure of separating the center-of-mass and relative coordinates as described above, we obtain the effective Hamiltonian for the relative mode as discussed in Appendix~\ref{sec:new addition}, by expanding the interaction for $|z| \ll R_0$:
\[
\begin{aligned}
H_{\mathrm{eff}}^{(1)}
={}&\frac{p^2}{m}+\frac{m\omega^2z^2}{4}
+\frac{C_6}{R_0^6}
-\frac{3C_6}{R_0^8}z^2\\
&+O(z^4).
\end{aligned}
\]
In this configuration, the interaction expansion contains no term
linear in the relative axial displacement $z$. As a result, the
interaction generates no state-dependent phase-space displacement
that requires cancellation. The leading contribution to the
controlled phase comes from the constant interaction energy. To obtain the phase, we neglect the quadratic and higher-order corrections in deriving the evolution operator, while accounting for their effects in the fidelity analysis. The resulting leading-order evolution operator is
\[
\hat V_1(t)=
e^{-i\omega t\hat a^\dagger\hat a}
\exp\left(-i\frac{C_6t}{\hbar R_0^6}\right).
\]}

\subsection{Implementation of the logical CZ gate}

We implement the logical CZ gate using two approaches, employing the evolution operators 
$\hat{V}(t)$ and $\hat{V_1}(t)$, respectively. The procedure begins by independently entangling
the motional GKP qubit of each atom with its corresponding ancilla, following the ancilla–motional entanglement procedure illustrated  in  Panel 1 of \figref{2}{}, preparing the two atoms in the joint state $|\Psi_1\rangle\otimes|\Psi_2\rangle$ where $|\Psi_1\rangle = \alpha|0_L^\Delta\rangle\otimes|0\rangle + \beta|1_L^\Delta\rangle\otimes|1\rangle$, $|\Psi_2\rangle = \gamma|0_L^\Delta\rangle\otimes|0\rangle + \delta|1_L^\Delta\rangle\otimes|1\rangle.$

The second stage, corresponding to phase generation, consists of two cycles for the approach using $\hat{V}(t)$ and single cycle using $\hat{V}_1(t)$. Following the preparation of the joint state, the first cycle begins with free harmonic evolution for a duration
$\pi/\omega-t/2$. A resonant $\pi$-pulse is then applied to each atom,
coupling the $|1\rangle$ ancilla state to the transient Rydberg state $| r\rangle$
while leaving \rev{the} $|0\rangle$ state uncoupled. Thus, only the
$|1_L1_L\rangle|11\rangle$ component of the joint state is doubly excited to
$|1_L1_L\rangle| rr\rangle$. This component subsequently evolves
under Rydberg interaction for a duration $t$. A second resonant $\pi$-pulse takes
$| rr\rangle$ back to $|11\rangle$, after which the system undergoes
another free-evolution interval of duration $\pi/\omega-t/2$ .

\begin{figure*}[t]
\hypertarget{figtop:6}{}
\centering
\includegraphics[width=\linewidth]{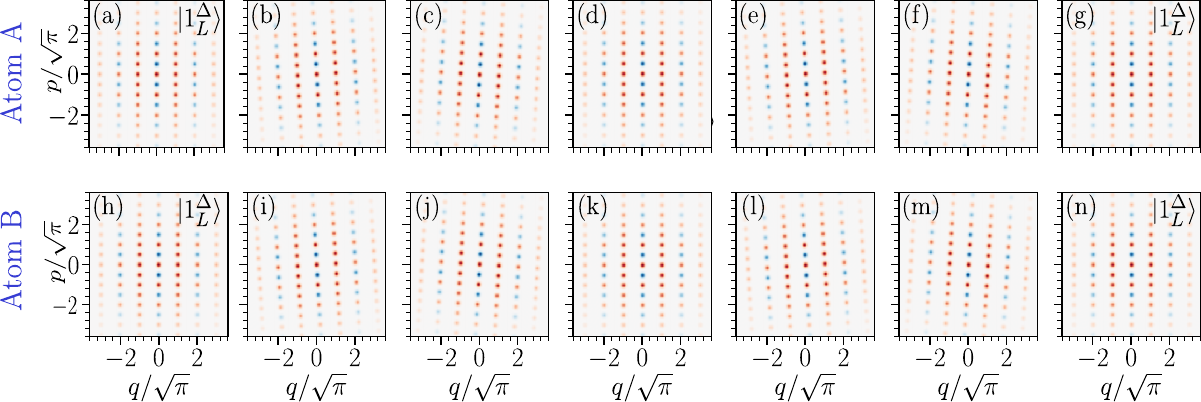}
\caption{\rev{Wigner functions of the two-atom logical state $|1_L^\Delta\rangle|1_L^\Delta\rangle$ during the second stage of the CZ gate in the adjacent-lattice configuration. The first and second rows show the evolution of atoms~A and B, respectively. Subplots~(a)--(d) and (h)--(k) illustrate the first cycle. The initial states evolve under the harmonic rotation $\hat{R}(\pi/\omega - t/2)$, giving states~(b) and (i), followed by the Rydberg interaction $\hat{V}(t)$, yielding~(c) and (j), and a subsequent harmonic rotation $\hat{R}(\pi/\omega - t/2)$, resulting in subplots~(d) and (k). Since the interaction acts on the relative coordinate, the small residual displacements at the end of the first cycle in (d) and (k) are equal and opposite for atoms A and B (see Appendix~\ref{sec:level19}) The second cycle, shown in subplots~(e)--(g) and (l)--(n), follows an analogous sequence, with the harmonic rotations replaced by $\hat{R}(2\pi/\omega - t/2)$. For $R_0 = 14.123\,\mu\mathrm{m}$ and $t = 0.564\,\mu\mathrm{s}$, the small residual displacements after the first cycle are cancelled by the second cycle, restoring  $|1_L^\Delta\rangle|1_L^\Delta\rangle$ to the initial motional configuration of both atoms.}}
\label{fig:6}
\end{figure*}

\rev{For the approach employing $\hat{V}(t)$, the motional evolution of the $|1_\Delta^L 1_\Delta^L\rangle$ component as shown in \figref{6}{} during the first cycle is}
\begin{equation}
e^{-i(\pi-\omega t/2)\hat{a}^{\dagger}\hat{a}}
\hat{V}(t)
e^{-i(\pi-\omega t/2)\hat{a}^{\dagger}\hat{a}}.
\end{equation}
The remaining logical components in the joint state do not excite their
corresponding ancillas to the doubly excited Rydberg state $| rr\rangle$
and therefore do not experience the Rydberg interaction, evolving instead only
under the free harmonic Hamiltonian during the interaction interval.
\begin{equation}
\begin{aligned}
e^{-i(\pi-\omega t/2)\hat{a}^{\dagger}\hat{a}}
e^{-i\omega t\hat{a}^{\dagger}\hat{a}}
e^{-i(\pi-\omega t/2)\hat{a}^{\dagger}\hat{a}}\\
=
e^{-i2\pi\hat{a}^{\dagger}\hat{a}}
=
\mathbb{I}.
\end{aligned}
\end{equation}
Using the expression for $\hat{V}(t)$, the evolution of the
$|1_L1_L\rangle$ component evaluates to 
\begin{equation}
\hat{D}\!\left(-2i\beta_d\sin\frac{\omega t}{2}\right)
\exp\!\left[
\frac{i\beta_d^{2}}{2}
\left(\omega t-\sin\omega t\right)
-\frac{iC_{6}t}{R_{0}^{6}\hbar}
\right].
\nonumber
\end{equation}
In the second cycle, the same sequence of free evolution, resonant
$\pi$-pulse excitation, Rydberg interaction, $\pi$-pulse de-excitation, and
free evolution is repeated, with the free-evolution intervals increased to
$2\pi/\omega-t/2$. The $|1_L1_L\rangle$ component therefore undergoes
\begin{equation}
e^{-i(2\pi-\omega t/2)\hat{a}^{\dagger}\hat{a}}
\hat{V}(t)
e^{-i(2\pi-\omega t/2)\hat{a}^{\dagger}\hat{a}},
\end{equation}
which produces a displacement of the same magnitude but opposite sign while
accumulating the same phase (see Appendix~\ref{sec:level19}).
As a result, after the application of these two cycles, the two displacements cancel exactly in phase space, as illustrated in \figref{6}{}, leaving only the accumulated phase after a total evolution time of $6\pi/\omega$. As the total evolution time is  a multiple of \(2\pi/\omega\), the evolution generated by $H_{\mathrm{com}}$ becomes the identity operator. It
therefore contributes no residual motional evolution to the logical
gate. The logical component $|1_L^\Delta1_L^\Delta\rangle$ acquires the phase factor $\exp\!\left[
i\,\beta_d^{2}
\bigl(\omega t - \sin(\omega t)\bigr)-i\frac{2C_6 t}{R_0^6\hbar}
\right]$ while all logical components return to their initial configuration with no residual displacement. By choosing the Rydberg 
interaction time $t$ such that the phase equals $\pi$, the protocol realizes a logical CZ gate conditioned on the simultaneous Rydberg excitation of both ancillas.

\rev{Alternatively, when using the evolution operator $\hat{V}_1(t_1)$, no displacement cancellation is required because $\hat{V}_1(t_1)$ contains no interaction-induced displacement. Therefore, the logical CZ gate can be implemented using only the first cycle, without requiring the second cycle.
\begin{equation}
\begin{aligned}
e^{-i(\pi - \omega t_1/2)\hat{a}^\dagger\hat{a}} \hat{V}_1(t_1) e^{-i(\pi - \omega t_1/2)\hat{a}^\dagger\hat{a}} \\
= \exp\left(-i\frac{C_6 t_1}{\hbar R_0^6}\right) e^{-i2\pi\hat{a}^\dagger\hat{a}}\\ 
= \exp\left(-i\frac{C_6 t_1}{\hbar R_0^6}\right) \hat{I}.
\end{aligned}   
\end{equation}
Hence, the motional state returns to its initial configuration after a single cycle, while the $|1_\Delta^L 1_\Delta^L\rangle$ component acquires the interaction phase factor $\exp[-iC_6 t_1/(\hbar R_0^6)]$. The interaction time $t_1$ is chosen such that this phase equals $\pi$, thereby implementing the logical CZ gate.}

In both approaches, the remaining logical components do not experience the
doubly excited Rydberg interaction and undergo only free harmonic evolution
during the interaction interval. Finally, the ancilla of each atom is independently disentangled from its corresponding motional state using the disentangling procedure illustrated in Panel 3 of \figref{2}{}. The resulting state is $(\alpha\gamma|0_L^{\Delta}\rangle_1|0_L^{\Delta}\rangle_2
+\alpha\delta|0_L^{\Delta}\rangle_1|1_L^{\Delta}\rangle_2
+\beta\gamma|1_L^{\Delta}\rangle_1|0_L^{\Delta}\rangle_2
-\beta\delta|1_L^{\Delta}\rangle_1|1_L^{\Delta}\rangle_2\bigr)\otimes|00\rangle$. Thus, the protocol realizes the desired logical CZ gate while returning both ancillas to their initial states.

\section{\label{sec:level13}Physical Realization of Gates}

As a representative physical implementation of the proposed single-qubit phase gate and the  CZ gate, we consider a neutral-atom optical lattice platform based on $^{88}$Sr atoms. The ancilla states $|0\rangle$, $|1\rangle$ correspond to $5s5p\,{}^3P_0$ and $5s5p\,{}^3P_2$, respectively, while $5s78\,{}^3S_1$ serves as a transient Rydberg state. A triple-magic trapping wavelength near $\lambda = 596~\mathrm{nm}$ has been proposed for this level configuration in Ref.~\cite{meinert2023state}, enabling nearly state-independent trapping of the ancilla and Rydberg states.
Large-scale optical lattice architectures for neutral atoms have already been demonstrated experimentally~\cite{tao2024prl}, providing a workhorse for the present proposal. The present scheme considers two parallel sparsely occupied two-dimensional optical-lattice planes, as shown in \figref{5}{(a)}, \rev{with the atom arranged so that the minimum interatomic separation between the simultaneously excited atoms is well beyond the Rydberg blockade radius.} The GKP qubit is encoded in the axial ($z$) motional mode of each lattice site, while the transverse ($x,y$) modes serve as spectator degrees of freedom. 

\rev{Entangling operations are implemented between atoms occupying corresponding lattice sites in opposite layers or between atoms in the same layer, depending on the chosen configuration. For atoms in opposite layers, the equilibrium interatomic separation is aligned with the encoded axial direction, whereas for atoms in the same layer, it lies in the xy-plane. In both configurations, we consider the relative axial displacement $z$ while assuming that the transverse relative displacements $x$ and $y$ remain unchanged during the interaction. The corresponding one-dimensional interaction potentials are therefore approximated as
\begin{equation}
V_{\mathrm{1D}}(z)=
\begin{cases}
\dfrac{C_6}{(R_0+z)^6}, & \text{opposite layers},\\[8pt]
\dfrac{C_6}{(R_0^2+z^2)^3}, & \text{same layer}.
\end{cases}
\end{equation}}
The validity of this one-dimensional description depends on the transverse spectator modes remaining effectively decoupled from the Rydberg interaction. To assess this condition, we employ the anharmonic treatment of Ref.~\cite{bohnmann2025bosonic}, which provides analytical expressions for the trap frequencies, anharmonicity, and leading-order transverse--axial mode-coupling coefficients. Based on this treatment, we choose a representative trap depth of $U_0/k_{\mathrm B}=6\,\mathrm{mK}$ and a beam waist of $w_0=6\,\mu\mathrm{m}$. These parameters yield an axial trap frequency of $\omega=\omega_z=2\pi\times40\,\mathrm{kHz}$ and transverse trap frequencies of $\omega_{x,y}=2\pi\times1.26\,\mathrm{MHz}$, while maintaining weak anharmonicity and small transverse--axial mode couplings (see Appendix~\ref{sec:level20}). To quantify the validity of neglecting the transverse displacements in the interaction potential, we compare this one-dimensional approximation with the full three-dimensional Rydberg interaction, accounting for the leading-order corrections due to $x$ and $y$ as presented in Appendix~\ref{sec:level21}. The deviation is negligible for the chosen parameters, confirming that the influence of the transverse motion on the interaction strength is insignificant.

In addition, the optical power required for a Gaussian optical lattice scales as $P\propto U_0w_0^2/\alpha$, where $\alpha$ is the dynamic polarizability of the trapped atomic state at the trapping wavelength. For the chosen trapping wavelength of $596~\mathrm{nm}$, \rev{and $\alpha\approx100$ [a.u.]},  the required optical power is comparable to that employed in the experimentally demonstrated large-scale optical lattice of Ref.~\cite{tao2024prl}, confirming the experimental feasibility of the proposed trapping configuration.

\begin{figure}[t]
    \hypertarget{figtop:7}{}
    \centering
    \includegraphics[width=0.48\textwidth]{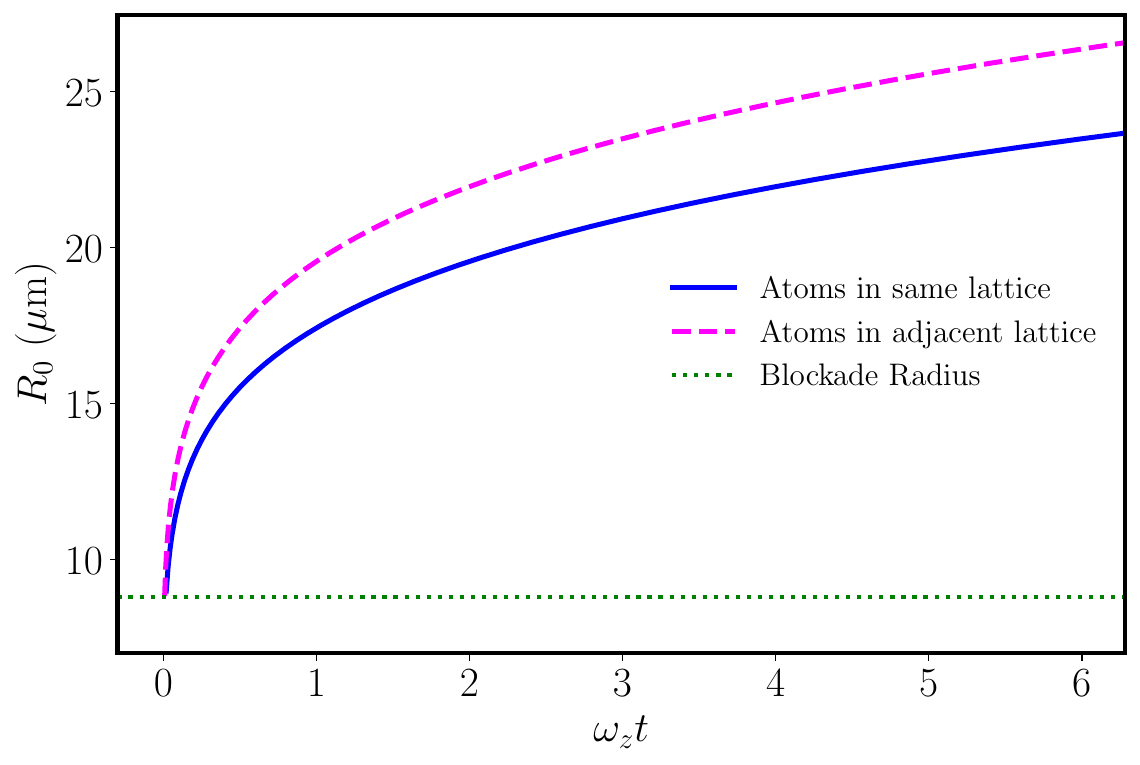}
    \caption{\rev{
Required interatomic separation $R_0$ for the CZ gate versus the dimensionless Rydberg interaction time $\omega_z t$, for atoms in the same lattice plane (solid curve) and in adjacent lattice planes (dashed curve). We estimate the blockade radius (dotted line) as $R_{\mathrm b}=8.8\,\mu\mathrm{m}$,
using $C_6$ for the $n=78$ Rydberg state and an assumed Rabi
frequency of $\Omega/2\pi=7.7\,\mathrm{MHz}$, based on the value
reported for $n=61$ in Ref.~\cite{PRXQuantum.6.010331}
In both configurations the required separation $R_0$ exceeds $R_b$, so the protocol operates in the weak-interaction regime.}}
    \label{fig:7}
\end{figure}

\rev{For the Rydberg state $5s78\,{}^{3}S_{1}$, we calculated the
interaction coefficient $C_6 \approx -2.33\times10^{-57}\,
\mathrm{J\,m^6}$ using ARC python library~\cite{Robertson2021ARC}.The CZ gate parameters are determined by imposing the phase-matching condition for the two approaches. For the two-cycle approach based on $\hat{V}(t)$, the accumulated phase must satisfy
\begin{equation}
\beta_d^2\bigl(\omega_z t-\sin(\omega_z t)\bigr)
-\frac{2C_6t}{\hbar R_0^6}
=\pi.
\label{eq:PhaseCondition}
\end{equation}
For the single-cycle approach based on $\hat{V}_1(t)$, which neglects the displacement contribution, the phase-matching condition reduces to
\begin{equation}
-\frac{C_6t}{\hbar R_0^6}=\pi.
\label{eq:PhaseCondition1}
\end{equation}
For the representative axial trap frequency $\omega_z=2\pi\times40\,\mathrm{kHz}$, the phase-matching conditions in Eqs.~\ref{eq:PhaseCondition} and~\ref{eq:PhaseCondition1} determine the required Rydberg interaction durations as functions of the interatomic separation $R_0$. The corresponding solutions for both equations are plotted in \figref{7}{}, allowing direct comparison of the two approaches. The obtained separations extend up to approximately $27\,\mu\mathrm{m}$ for adjacent-plane atoms, and $22\,\mu\mathrm{m}$ approximately for in-plane atoms, both well outside the blockade radius, providing experimentally realistic operating points where simultaneous excitation of both atoms to the Rydberg state is feasible while retaining a finite interaction-induced phase. 

The two configurations offer complementary advantages. The adjacent-plane configuration provides greater flexibility in controlling the interatomic separation, as the distance between the lattice planes can be independently adjusted. The in-plane configuration, on the other hand, avoids the additional interaction cycle required for displacement cancellation, resulting in a faster gate with less motional squeezing error. Furthermore, the adjacent-layer configuration, together with in-plane interactions, can provide up to six nearest-neighbour connections per atom, offering a connectivity compatible with surface-code architectures}

The parameters and their values are considered in such a way that the proposed protocol can be realized with current experimental capabilities. A more detailed optimization of the trapping parameters remains an important direction for future work. Nevertheless, the representative parameter set considered here indicates that the proposed CZ gate can be implemented using experimentally realistic trapping frequencies, Rydberg interaction strengths, and interatomic separations available in current neutral-atom platforms.

\section{\label{sec:level14}Gate Timing and Fidelity}

The duration of the proposed logical gates is determined primarily by the motional evolution of the harmonic oscillator. Throughout this work, we assume that operations acting solely on the internal ancilla states can be performed on a timescale much shorter than that of the motional dynamics~\cite{unnikrishnan2024coherent} and therefore make a negligible contribution to the total gate duration.

In the protocol considered here, the logical Hadamard gate requires a motional evolution time of $\pi/(2\omega)$, whereas both the single-qubit phase and adjacent plane CZ gate comprise three sequential stages: motional--ancilla entanglement for a duration $2\pi/\omega$, phase generation for $6\pi/\omega$, and motional--ancilla disentanglement for $4\pi/\omega$, giving a total gate duration of $12\pi/\omega$.
At $\omega=2\pi\times40~\mathrm{kHz}$, the corresponding gate durations are $6.25~\mu\mathrm{s}$ for the Hadamard gate and $150~\mu\mathrm{s}$ for both the single-qubit phase and adjacent plane CZ gate. \rev{The obtained gate durations are well below the estimated GKP state survival time in our parameter regime. In Ref.~\cite{bohnmann2025bosonic} the characteristic GKP state survival time is estimated to be approximately 20 oscillator cycles for the optical lattice parameters considered therein. For our parameter regime, the axial anharmonicity is approximately 1.7 times larger (see Appendix~\ref{sec:level20}), resulting in a reduced estimated survival time of approximately 17 oscillator cycles. This limitation can be mitigated by employing potential painting technique~\cite{Henderson_2009} to eliminate the fourth-order axial anharmonicity, effectively setting $\eta_z=0$~\cite{Pampel_2026}. In addition, the transverse couplings $\epsilon_{zy}$ and $\epsilon_{zx}$ are also suppressed. With these modifications, the estimated GKP state survival time can be extended to approximately 50 oscillator cycles, although this comes at the cost of longer physical gate durations due to the associated reduction in axial and transverse trap frequencies.}

\subsection{\label{sec:level15}Single qubit phase gate fidelity}

To quantify the performance of the proposed logical phase gate, we evaluate the fidelity between the oscillator state produced by the protocol and the corresponding ideal logical target state. \rev{Since the ancilla is not measured during the protocol, the GKP state is obtained by tracing out the ancilla state,
\begin{equation}
\rho_{\mathrm{osc}} = \mathrm{Tr}_{\mathrm{anc}} \left( |\Psi_{\mathrm{out}}\rangle\langle\Psi_{\mathrm{out}}| \right),
\end{equation}
where $|\Psi_{\mathrm{out}}\rangle$ denotes the final joint motional--ancilla state. An arbitrary pure logical input state is parameterized as
$
|\psi_{\mathrm{in}}\rangle = \cos\frac{\vartheta}{2}|0_\Delta^L\rangle + e^{i\varphi}\sin\frac{\vartheta}{2}|1_\Delta^L\rangle,
$
where $\vartheta\in[0,\pi]$ and $\varphi\in[0,2\pi)$. The ideal action of the logical phase gate on this input state gives
$
|\psi_{\mathrm{target}}\rangle = \cos\frac{\vartheta}{2}|0_\Delta^L\rangle + e^{i\varphi}e^{i\theta_{\mathrm{g}}}\sin\frac{\vartheta}{2}|1_\Delta^L\rangle,
$
where $\theta_{\mathrm{g}}$ denotes the phase imparted by the logical phase gate. The state fidelity is computed as
$
F(\vartheta,\varphi) = \langle\psi_{\mathrm{target}}| \rho_{\mathrm{osc}} |\psi_{\mathrm{target}}\rangle.
$
The average gate fidelity ($\overline{F}$) is obtained by averaging the state fidelity over the Haar-uniform distribution of pure states in the two-dimensional logical Hilbert space,
\begin{equation}
\overline{F} = \frac{1}{4\pi} \int_0^\pi\int_0^{2\pi} F(\vartheta,\varphi)\sin\vartheta\, d\varphi\, d\vartheta.
\end{equation}}

The protocol requires neither ancilla measurement nor feedforward. This is particularly advantageous for motional GKP qubits, since fluorescence detection imparts recoil to the trapped atom and can disturb the encoded motional state. The protocol is designed to return the ancilla deterministically to $|0\rangle$, leaving the oscillator in the desired logical state.

\begin{figure}[t]
    \hypertarget{figtop:8}{}
    \includegraphics[width=0.50\textwidth]{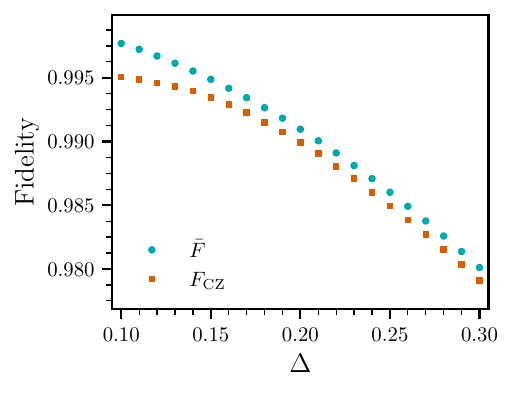}
    \caption{Average logical gate fidelities $\bar{F}$ and $F_{\mathrm{cz}}$ as functions of the finite-energy parameter $\Delta$ for the logical phase gate and the logical controlled-$Z$ gate between atoms in adjacent lattice planes, respectively. In both cases, the fidelity decreases with increasing $\Delta$ due to the reduced quality of the finite-energy GKP codewords, while remaining above $0.98$ over the range considered.}
    \label{fig:8}
\end{figure}

For ideal GKP states, the logical codewords are exact eigenstates of the logical $Z$ displacement used in the sequence, ensuring perfect disentanglement between the oscillator and ancilla. For finite-energy GKP states, however, the disentangling operation is imperfect, leaving residual oscillator--ancilla entanglement. Tracing out the ancilla therefore yields a mixed oscillator state. As $\Delta$ decreases, the finite-energy states approach the ideal codewords, reducing this residual entanglement and improving the gate fidelity. Hence, the average logical gate fidelity $\bar{F}$ increases monotonically with decreasing $\Delta$, as shown in \figref{8}{}.

\subsection{\label{sec:level16}Controlled-Z gate fidelity}

\rev{To characterize the performance of the proposed logical CZ gate, we evaluate the fidelity between the motional state obtained at the end of the protocol and the corresponding ideal target logical state. The reduced state of the motional modes is obtained by tracing over the ancillas,
$
\rho_{\mathrm{cz}}=\mathrm{Tr}_{\mathrm{anc}}
\left(|\Psi_{\mathrm{cz}}\rangle\langle\Psi_{\mathrm{cz}}|\right),
$
where \(|\Psi_{\mathrm{cz}}\rangle\) denotes the final composite oscillator–ancilla state. An arbitrary pure logical two-qubit input state is parameterized as
$
|\psi_{\mathrm{in}}\rangle={}
\alpha_1|0_\Delta^L0_\Delta^L\rangle
+\alpha_2|0_\Delta^L1_\Delta^L\rangle
+\alpha_3|1_\Delta^L0_\Delta^L\rangle
+\alpha_4|1_\Delta^L1_\Delta^L\rangle,
$
where \(\sum_{j=1}^{4}|\alpha_j|^2=1\). The ideal logical CZ gate applies a phase factor of \(-1\) only to the \(|1_\Delta^L1_\Delta^L\rangle\) component, yielding the target state
$
|\psi_{\mathrm{target}}\rangle={}
\alpha_1|0_\Delta^L0_\Delta^L\rangle
+\alpha_2|0_\Delta^L1_\Delta^L\rangle
+\alpha_3|1_\Delta^L0_\Delta^L\rangle
-\alpha_4|1_\Delta^L1_\Delta^L\rangle.
$
The state fidelity is computed as
$
F_1=\langle\psi_{\mathrm{target}}|
\rho_{\mathrm{cz}}|\psi_{\mathrm{target}}\rangle.
$
The average CZ gate fidelity \(F_{\mathrm{cz}}\) is obtained by averaging the state fidelity over the Haar-uniform distribution of pure states in the two-qubit logical Hilbert space.}

One source of CZ gate infidelity is imperfect
ancilla--motional disentanglement, which produces an error analogous
to that encountered in the single-qubit phase gate. \rev{In addition, the Rydberg interaction introduces a residual quadratic contribution to the motional dynamics, resulting in displacement and squeezing errors in phase space. The form of this contribution depends on the lattice configuration: it is $21 C_6 z^2 / R_0^8$ for atoms in adjacent lattices and $3 C_6 z^2 / R_0^8$ for atoms in the same lattice. Thus, although motional distortion occurs in both configurations, its magnitude is seven times larger for atoms in  adjacent planes at same $R_0$. We therefore focus on the adjacent-lattice configuration in the following fidelity analysis, as it represents the more stringent case for assessing the effect of motional distortion.}

Since the coefficient decreases as $R_0^{-8}$, increasing the interatomic separation suppresses this contribution. We consider the range  $R_0\simeq13.5$--$27~\mu\mathrm{m}$, which lies outside the Rydberg blockade radius and provides sufficient separation from other nearby atoms to minimize unwanted Rydberg interactions. Over this range, the calculated fidelity varies by less than $7.7\times10^{-4}$, indicating that the gate performance is only weakly dependent on the choice of $R_0$. At larger
$R_0$, however, the gate protocol requires longer Rydberg excitation time, making the gate more susceptible to control errors. We therefore choose $R_0=14.123~\mu\mathrm{m}$ \rev{as a representative operating point, balancing the suppression of motional distortion against the duration of the Rydberg interaction.} This choice corresponds to $\omega_z t=0.1418$. At $\omega_z = 2\pi\times40\,\mathrm{kHz}$, each Rydberg interaction interval lasts $0.564\,\mu\mathrm{s}$, resulting in a total Rydberg interaction time of approximately $1.128\,\mu\mathrm{s}$ for the two cycles in the protocol. This timescale is much shorter than the typical
Rydberg-state lifetime, making spontaneous decay during the interaction
a negligible contribution to the overall gate infidelity.

We next examine the full CZ-gate fidelity as a function of the
finite-energy parameter $\Delta$, including both the residual
ancilla--motional entanglement and the motional distortion arising from
the quadratic interaction. As $\Delta$ is decreased, the finite-energy
GKP states approach the ideal codewords, reducing the residual
ancilla--motional entanglement and thereby improving the gate fidelity.
The improvement, however, becomes progressively weaker at smaller
$\Delta$, as shown
in \figref{8}{}, indicating the emergence of an additional limitation.

\begin{figure}[t]
    \hypertarget{figtop:9}{}
    \raggedright
    \includegraphics[width=8cm,height=5.6cm]{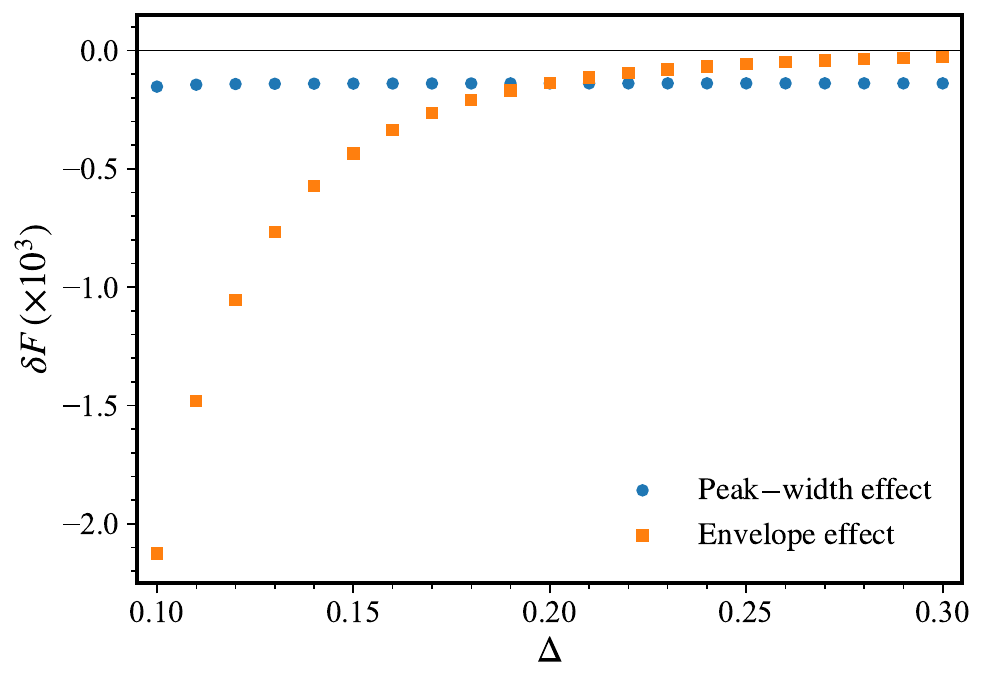}
    \caption{Change in the average logical CZ-gate fidelity, $\delta F$ for the adjacent lattice configuration, defined as the fidelity obtained with the quadratic error term minus the fidelity obtained without it, as a function of the finite-energy parameter $\Delta$. This illustrates how the error affects the gate fidelity through the two different $\Delta$-dependent features of the finite-energy GKP state.}
    \label{fig:9}
\end{figure}

To clarify the origin of this behavior, we isolate the effect of the residual quadratic interaction. The parameter $\Delta$ influences both the width of the individual GKP peaks and the finite-energy envelope. Decreasing $\Delta$ narrows the individual peaks, thereby reducing their sensitivity to the quadratic interaction. At the same time, however, it broadens the finite-energy envelope through the coefficients $e^{-2\pi\Delta^2 k^2}$, increasing the weight of peaks at larger $|z|$. These outer peaks are more strongly affected by the motional distortion, since the quadratic contribution scales as $z^2$. As shown in \figref{9}{}, both contributions decrease the fidelity as $\Delta$ is reduced; however, the broadened finite-energy envelope produces a considerable reduction in fidelity at small $\Delta,$ leading to the observed gradual saturation in fidelity. Further improvement would therefore require suppression of the residual quadratic interaction rather than a further reduction of $\Delta$.

\section{\label{sec:level15}Conclusion And Outlook}
In this paper, we have presented a universal set of logical gates for finite-energy GKP qubits encoded in the motional states of neutral atoms confined within an optical lattice. We proposed a deterministic protocol using harmonic evolution, ancilla-mediated conditional displacements, and weak Rydberg interactions to realize the gates. We outlined how two
sparsely occupied two-dimensional optical lattice planes separated along
the axial direction provide a suitable geometry for implementing the
proposed gate operations. Strong transverse confinement suppresses in-plane motion, allowing the Rydberg-mediated interaction to be reduced to an effective one-dimensional axial potential, which can be expanded around the equilibrium layer separation to obtain the state-dependent terms required for the gate dynamics. For $^{88}\mathrm{Sr}$, with
$\omega_z=2\pi\times40\,\mathrm{kHz}$, the Hadamard gate takes
 $6.25\,\mu\mathrm{s}$, while the single-qubit phase and the adjacent plane CZ gates require
$150\,\mu\mathrm{s}$. The same-plane CZ gate requires $100\,\mu\mathrm{s}$. We find that the dominant intrinsic errors arise from residual
ancilla--motion entanglement and the quadratic Rydberg-interaction term,
which induces unwanted displacement and squeezing.

 A comprehensive open-system analysis incorporating decoherence mechanisms, such as photon-scattering-induced recoil, laser phase and intensity noise, and trap-frequency fluctuations, would provide a more complete assessment of gate performance under realistic operating conditions. Further optimization of the trapping geometry and interaction parameters may lead to even higher gate fidelities and shorter operation times. Moreover, integrating the universal gate set proposed here with recently developed neutral-atom GKP state preparation and error-correction protocols, as well as concatenated surface-GKP architectures, represents an important step toward scalable fault-tolerant quantum computation based on motional bosonic qubits.

\section*{ACKNOWLEDGMENTS}
This work is supported by the Department of Science and Technology (DST), Government of India, under the National Quantum Mission (NQM). A.K., A.N.R., D.T., and I.K.D. acknowledge financial support from this mission.

\section*{Data Availability}
The data and code supporting the findings of this article are not publicly available because they are part of ongoing work. They are available from the authors upon reasonable request.

\appendix
\section{\label{sec:level17}Hadamard gate implementation}

The logical Hadamard gate for the square GKP code is implemented by
a phase-space rotation of angle \( \pi/2 \),
\begin{equation}
\hat H_L
=
e^{-i\frac{\pi}{2}\hat a^\dagger \hat a}.
\label{eq:hadamard}
\end{equation}
Since both the finite-energy operator
\(e^{-\Delta^2 \hat a^\dagger \hat a}\)
and the logical Hadamard operator
\(e^{-i\frac{\pi}{2}\hat a^\dagger \hat a}\)
are functions solely of the number operator
\(\hat a^\dagger \hat a\),
they commute 

\begin{equation}
\left[
e^{-i\frac{\pi}{2}\hat a^\dagger \hat a},
e^{-\Delta^2 \hat a^\dagger \hat a}
\right]
=0.
\label{eq:commutation}
\end{equation}
Consequently, the action of the Hadamard gate on a finite-energy GKP
state becomes

\begin{align}
\hat H_L |\psi^\Delta_L\rangle
&=
2\sqrt{\pi}\,\Delta\,
\hat H_L
e^{-\Delta^2 \hat a^\dagger \hat a}
|\psi_L\rangle
\\[4pt]
&=
2\sqrt{\pi}\,\Delta\,
e^{-\Delta^2 \hat a^\dagger \hat a}
\hat H_L
|\psi_L\rangle .
\label{eq:hadamard_action}
\end{align}
Therefore, the logical Hadamard
gate is implemented exactly even in the presence of the
finite-energy suppression factor.

\section{\label{sec:level18}Conditional displacement}

The conditional-displacement protocol employed in this work follows Ref.~\cite{bohnmann2025bosonic}. A state-dependent force is generated by an additional Gaussian laser beam operating at the tune-out wavelength. The beam has waist $w_{1}$ and trap depth $U_{1}$, and is displaced by a distance $w_{1}/2$ from the trap center. At the trap center, the curvature of this Gaussian potential vanishes so that the harmonic confinement is neither strengthened nor weakened. Consequently, the atom experiences a nearly constant force
$
f={2e^{-1/2}U_{1}}/{w_{1}}.
$
The motional dynamics are then governed by the Hamiltonian
\begin{equation}
\hat{H}
=
\frac{\hat{p}^{2}}{2m}
+
\frac{1}{2}m\omega^{2}\hat{x}^{2}
-
f\hat{x},
\label{eq:B1}
\end{equation}
where $m$ is the atomic mass and $\omega$ is the trap frequency.
\rev{Following the derivation in Ref.~\cite{bohnmann2025bosonic}, an effective momentum displacement is obtained by allowing the oscillator to evolve freely for a duration $\pi/\omega - t/2$, applying the force for a duration $t$, and then allowing the oscillator to evolve freely for another interval of duration $\pi/\omega - t/2$. Similarly, choosing the free-evolution duration as $2\pi/\omega - t/2$ produces a displacement in the opposite direction.

Accordingly, we define the conditional-displacement operators as
\begin{equation}
\hat{D}\left[-2i\alpha_d\sin\left(\frac{\omega t}{2}\right)\right] e^{i\theta(t)},
\end{equation}
and
\begin{equation}
\hat{D}\left[2i\alpha_d\sin\left(\frac{\omega t}{2}\right)\right] e^{i\theta(t)},
\end{equation}
where
\begin{equation}
\theta(t) = \frac{\alpha_d^2}{2} \left[\omega t - \sin(\omega t)\right].
\end{equation}
Here, $\alpha_d = f/\sqrt{\hbar m\omega^3}$, with $f$ denoting the magnitude of the applied force.}
In the simulations presented in the main text, the tune-out laser is assumed to have a trap depth $U_1 = 0.07\,U_0$. Using the optical lattice parameters summarized in Appendix~\ref{sec:level20}, the resulting displacement amplitude is
$
\alpha_d = 2.37.$
Since the force is conditioned on the internal ancilla state, the resulting displacement is likewise state dependent and forms the fundamental building block for the logical gate protocols presented in the main text.

\section{\label{sec:level19}CZ gate mechanism for atoms in adjacent lattice planes}
We consider two atoms trapped in similar harmonic potentials separated by a distance $R_0$ and interacting via  the Rydberg interaction. The total Hamiltonian is

\begin{equation}
\begin{aligned}
H =& \frac{p_1^2}{2m} + \frac{1}{2} m \omega^2 z_1^2 
+ \frac{p_2^2}{2m} + \frac{1}{2} m \omega^2 z_2^2 \\
&+ \frac{C_6}{\left| R_0 + (z_1 - z_2) \right|^6}.
\end{aligned}
\end{equation}
The relative displacement is defined by
\begin{equation}
z = z_1 - z_2,
\end{equation} assuming that $|z| \ll R_0$.
The interaction potential 
can then be expanded using a binomial expansion 
\begin{equation}
\frac{C_6}{|R_0 + z|^6}
\approx \frac{C_6}{R_0^6}
- \frac{6C_6}{R_0^7} z
+ \frac{21C_6}{R_0^8} z^2.
\end{equation}
Next, we transform to center-of-mass and relative coordinates:
\begin{align}
Z &= \frac{z_1 + z_2}{2}, & P &= p_1 + p_2,\label{eqn:C4} \\
z &= z_1 - z_2, & p &= \frac{p_1 - p_2}{2}.
\label{eqn:C5}
\end{align}
Using these variables, the kinetic energy becomes
\begin{equation}
\frac{p_1^2}{2m} + \frac{p_2^2}{2m}
= \frac{P^2}{4m} + \frac{p^2}{m},
\end{equation}
and the harmonic potential transforms to
\begin{equation}
\frac{1}{2} m \omega^2 (z_1^2 + z_2^2)
= m \omega^2 Z^2 + \frac{1}{4} m \omega^2 z^2.
\end{equation}
Putting it all together, the resultant Hamiltonian separates into center-of-mass and relative parts:
\begin{equation}
H = H_{\mathrm{com}} + H_{\mathrm{rel}}.
\end{equation}
Thus, the total Hamiltonian can be written explicitly as
\begin{align}
H &= \frac{P^2}{4m} + m \omega^2 Z^2 \\
\nonumber
&+ \frac{p^2}{m} + \frac{1}{4} m \omega^2 z^2  \\
\nonumber
&+ \frac{C_6}{R_0^6}
- \frac{6C_6}{R_0^7} z
+ \frac{21C_6}{R_0^8} z^2.
\end{align}

The center-of-mass motion remains a simple harmonic oscillator, while the relative motion experiences both a modified quadratic confinement and a linear term due to the Rydberg interaction, indicating a shifted equilibrium position and an effective change in trapping frequency.

Since $H_{\mathrm{com}}$ depends only on $(Z, P)$ and $H_{\mathrm{rel}}$ depends only on $(z, p)$, it follows that
\begin{equation}
[H_{\mathrm{com}}, H_{\mathrm{rel}}] = 0.
\end{equation}
Therefore, the total Hamiltonian is separable, and the center-of-mass and relative motions evolve independently. Each can be treated as an independent harmonic oscillator with the relative mode modified by the interaction term.

For the relative motion, we retain only  the constant and linear terms in the interaction:
\begin{equation}
H_{\mathrm{eff}} \approx \frac{p^2}{m} + \frac{1}{4} m \omega^2 z^2 
+ \frac{C_6}{R_0^6} - \frac{6C_6}{R_0^7} z.
\end{equation}
In terms of the quadrature operators, this can be written as 
\begin{equation}
H_{\mathrm{eff}}=
\frac{\hbar\omega}{2}
\left[
\hat{p}^{\,2}
+
\left(\hat{q}-\beta_d\right)^2
\right]
-\frac{\beta_d^2\hbar\omega}{2}
+\frac{C_6}{R_0^6},
\end{equation}
where $
\beta_d=
\frac{6C_6}{R_0^7\,\sqrt{\hbar m\omega^3/2}}$, the evolution under this Hamiltonian for a time $t$ reduces to
unitary $\hat V(t)$ such that 
\begin{equation}
\begin{split}
\hat V(t)={}&\hat{D}\!\left[\beta_d\!\left(1-e^{-i\omega t}\right)\right]
e^{-i\omega t\,\hat{a}^\dagger\hat{a}}\\
&\times e^{\frac{i}{2}\beta_d^2\bigl(\omega t-\sin(\omega t)\bigr)-i\frac{C_6 t}{R_0^6\hbar}}
\end{split}
\end{equation}
Consider a duration of \(2\pi/\omega\) in which we start by leaving the oscillator idle for $\frac{\pi}{\omega}-\frac{t}{2}$, then applying the unitary $\hat V(t)$ and letting the oscillator idle again for a time $\frac{\pi}{\omega}-\frac{t}{2}$ which yields
\begin{align}
&e^{-i\left(\pi-\frac{\omega t}{2}\right)\hat a^\dagger \hat a}\,
\hat V(t)\,
e^{-i\left(\pi-\frac{\omega t}{2}\right)\hat a^\dagger \hat a}
\nonumber\\[6pt]
&= \hat D\!\left(-2i\beta_d\sin\frac{\omega t}{2}\right)\,
e^{\frac{i}{2}\beta_d^2\bigl(\omega t-\sin(\omega t)\bigr)-i\frac{C_6 t}{R_0^6\hbar}} .
\end{align}
Now consider a second evolution period of \(4\pi/\omega\), during
which the oscillator evolves freely for time $\frac{2\pi}{\omega}-\frac{t}{2}$, then applying the unitary $\hat V(t)$ again by using the tuned-out laser and then letting the oscillator idle again for $\frac{2\pi}{\omega}-\frac{t}{2}$ which yields 
\begin{align}
&e^{-i\left(2\pi-\frac{\omega t}{2}\right)\hat a^\dagger \hat a}\,
\hat V(t)\,
e^{-i\left(2\pi-\frac{\omega t}{2}\right)\hat a^\dagger \hat a}
\nonumber\\[6pt]
&= \hat D\!\left(2i\beta_d\sin\frac{\omega t}{2}\right)\,
e^{\frac{i}{2}\beta_d^2\bigl(\omega t-\sin(\omega t)\bigr)-i\frac{C_6 t}{R_0^6\hbar}} .
\end{align}
Applying these evolutions one after the other for a total time period of \(6\pi/\omega\), the displacements from the two time
intervals cancel exactly in phase space and the finite dynamical
phase $\exp\!\left[
i\,\beta_d^{2}
\bigl(\omega t - \sin(\omega t)\bigr)-i\frac{2C_6 t}{R_0^6\hbar}
\right]$ remains.

\rev{The above derivation is performed in the centre-of-mass and relative coordinates, whereas the logical GKP qubits are encoded in the individual atomic modes $z_1$ and $z_2$. The connection between the two descriptions is established by writing the linear term as
$-\frac{6C_6}{R_0^7}(z_1 - z_2) = -f z_1 + f z_2, f = \frac{6C_6}{R_0^7}.$
This corresponds to equal and opposite forces on the two atoms. Consequently, the relative-mode displacements in Eqs.~(C14) and (C15) translate into equal and opposite displacements of the individual atomic modes, each with amplitude
$\alpha_{\mathrm{vdW}} = \frac{f}{\sqrt{\hbar m\omega^3}} = \frac{\beta_d}{\sqrt{2}}.$
These displacements are cancelled on each atom after the second cycle, restoring the motional states to their initial phase-space configurations and leaving only the accumulated phase.
$\frac{21C_6}{R_0^8}(z_1^2 + z_2^2 - 2z_1z_2)$
The terms proportional to $z_1^2$ and $z_2^2$ produce independent single-atom squeezing errors, while the cross term,
$-\frac{42C_6}{R_0^8}z_1z_2,$
introduces a correlated interaction between the motional modes of the two GKP qubits, whose effect is already included in fidelity analysis.}

\section{\label{sec:new addition}CZ gate mechanism for atoms in the same lattice plane}
\rev{We now consider two atoms trapped in similar harmonic potentials and separated by a fixed transverse distance $R_0$, such that their equilibrium positions differ in the $xy$-plane, while their motional dynamics of interest remains along the $z$-axis. The interatomic separation is therefore $R=\sqrt{R_0^2+(z_1-z_2)^2}$, and the total Hamiltonian is given by
\begin{equation}
\begin{aligned}
H_1 ={} &\frac{p_1^2}{2m}+\frac{1}{2}m\omega^2z_1^2 + \frac{p_2^2}{2m}+\frac{1}{2}m\omega^2z_2^2 \\
&+\frac{C_6}{\left[R_0^2+(z_1-z_2)^2\right]^3}.
\end{aligned}
\end{equation}
Assuming that the relative axial displacement is small compared with the transverse separation, $|z_1-z_2|=|z|\ll R_0$, the Rydberg interaction can be expanded as
\begin{equation}
\begin{aligned}
\frac{C_6}{\left[R_0^2+(z_1-z_2)^2\right]^3}
&=\frac{C_6}{R_0^6}\left[1+\frac{(z_1-z_2)^2}{R_0^2}\right]^{-3} \\
&\approx \frac{C_6}{R_0^6} - \frac{3C_6}{R_0^8}z^2.
\end{aligned}
\end{equation}
Thus, the total Hamiltonian can be written explicitly as
\begin{align}
H &= \frac{P^2}{4m} + m \omega^2 Z^2 
+ \frac{p^2}{m} + \frac{1}{4} m \omega^2 z^2  \\
\nonumber
&+ \frac{C_6}{R_0^6}
- \frac{3C_6}{R_0^8} z^2.
\end{align}
where, $Z, P, z, p$ are same as defined in \eqref{eqn:C4}, \eqref{eqn:C5}.
By separating the center-of-mass and relative modes in the same manner as in Appendix~\ref{sec:level19}, we obtain the effective Hamiltonian for the relative mode:
\begin{equation}
H_{\mathrm{eff}}^{(1)} = \frac{p^2}{m} + \frac{m\omega^2}{4}z^2  + \frac{C_6}{R_0^6}- \frac{3C_6}{R_0^8}z^2.
\end{equation}
Unlike the case in which the atoms are separated along the axial direction only, the interaction expansion contains no term linear in the relative coordinate, so the interaction does not drive a phase-space displacement of the relative mode. Consequently, no displacement-cancellation sequences are required during the gate. Treating the quadratic term $-\frac{3C_6}{R_0^8}z^2$ as a motional squeezing error on the relative mode, the dominant interaction-induced phase is governed by the constant energy shift $C_6 / R_0^6$.
This constant energy shift $C_6 / R_0^6$ gives rise to a phase,
\begin{equation}
\phi_{\mathrm{int}} = -\frac{C_6 t}{\hbar R_0^6},
\end{equation}
accumulated over the Rydberg interaction time $t$. By tuning the interaction duration such that $\phi_{\mathrm{int}} = \pi$, this interaction directly implements a controlled-Z ($\mathrm{CZ}$) gate}.

\section{\label{sec:level20}Optical lattice parameters}

The trapping parameters used in the main text are obtained from the anharmonic treatment of a two-dimensional square optical lattice developed in Ref.~\cite{bohnmann2025bosonic}. For completeness, we summarize the expressions used to calculate the trap frequencies, anharmonicity, and transverse-axial mode-coupling coefficients for the representative lattice considered in this work. These expressions are evaluated for the chosen lattice depth, beam waist, and trapping wavelength to obtain the numerical values used in the simulations.

\begin{table}[h]
\centering
\caption{Representative optical lattice parameters used in this work.}
\label{tab:lattice_parameters}
\begin{tabular}{lc}
\hline\hline
Parameter & Value \\
\hline
Trapping wavelength $\lambda$ & $596~\mathrm{nm}$ \\
Trap depth $U_0/k_B$ & $6~\mathrm{mK}$ \\
Beam waist $w_0$ & $6~\mu\mathrm{m}$ \\
Axial trap frequency $\omega_z$ & $2\pi\times40~\mathrm{kHz}$ \\
Transverse trap frequency $\omega_{x,y}$ & $2\pi\times1.26~\mathrm{MHz}$ \\
Axial anharmonicity $\eta_z$ & $4\times10^{-5}$ \\
Mode coupling $\epsilon_{zx}=\epsilon_{zy}$ & $2.5\times10^{-3}$ \\
\hline\hline
\end{tabular}
\end{table}

For a two-dimensional square optical lattice, the harmonic trapping frequencies are given by
$\omega_z = \frac{2}{w_0}\sqrt{\frac{U_0}{m}},
\quad\omega_{x,y} = \frac{2\pi}{\lambda}\sqrt{\frac{U_0}{m}},$
where $U_0$ is the lattice depth, $w_0$ is the beam width, $\lambda$ is the trapping wavelength, and $m$ is the atomic mass. The leading-order anharmonicity of the axial mode is
$\eta_z=\frac{\hbar\omega_z}{8U_0}$,
while the transverse-axial mode-coupling coefficients are
$\epsilon_{zx}=\epsilon_{zy}
=\frac{\hbar\omega_{x,y}}{4U_0}.
$

The calculated lattice parameters are summarized in Table~\ref{tab:lattice_parameters}. As already discussed, the small values of $\eta_z$ and $\epsilon_{zx,zy}$ indicate that the axial motional mode is well described by the harmonic approximation, and is weakly coupled to the transverse spectator modes.

\section{\label{sec:level21}Error in the One-Dimensional Interaction Approximation}
To quantify the validity of the one-dimensional approximation employed in the main text for atoms in adjacent lattice planes, we compare the full three-dimensional Rydberg interaction with the interaction obtained by neglecting the transverse degrees of freedom. The interaction potential is
$
V=\frac{C_6}{R^6},
$
where the instantaneous interatomic separation is
$
R=\sqrt{(R_0+z)^2+x^2+y^2}.
$
Here, \(R_0\) is the equilibrium separation along the \(z\)-axis, \(z\) is the relative axial displacement, and \(x\) and \(y\) are the relative transverse displacements. Throughout the main text, the interaction is approximated by
$
V_{\mathrm{1D}}
=
\frac{C_6}{(R_0+z)^6},
$
which neglects the transverse coordinates.
Defining
$
r^2=x^2+y^2,
$
the interatomic separation can be written as
\begin{equation}
R
=
(R_0+z)
\sqrt{1+\frac{r^2}{(R_0+z)^2}},
\end{equation}
such that
\begin{equation}
V
=
\frac{C_6}{(R_0+z)^6}
\left(
1+\frac{r^2}{(R_0+z)^2}
\right)^{-3}.
\end{equation}
Using the binomial expansion gives
\begin{equation}
V
\simeq
\frac{C_6}{(R_0+z)^6}
\left[
1
-
3\frac{x^2+y^2}{(R_0+z)^2}
+\cdots
\right].
\end{equation}
Therefore, the leading-order fractional error introduced by neglecting the transverse motion is
\begin{equation}
\frac{\Delta V}{V_{1D}}
=
\frac{|V-V_{\mathrm{1D}}|}{V_{\mathrm{1D}}}=
3
\frac{x^2+y^2}{(R_0+z)^2}.
\label{eq:dV}
\end{equation}
For atoms each prepared in its transverse motional ground state,  the relative displacement variances are
$
\langle x^2\rangle=\frac{\hbar}{m\omega_x},
\langle y^2\rangle=\frac{\hbar}{m\omega_y},
$
such that
$
\left\langle x^2+y^2\right\rangle
=
\frac{\hbar}{m}
\left(
\frac{1}{\omega_x}
+
\frac{1}{\omega_y}
\right).$ The average fractional error therefore becomes
\begin{equation}
\left\langle\frac{\Delta V}{V}\right\rangle
=
\frac{3\hbar}{m(R_0+z)^2}
\left(
\frac{1}{\omega_x}
+
\frac{1}{\omega_y}
\right).
\end{equation}
For the parameters used in this work,
$
R_0=14.123~\mu\mathrm{m},
\omega_x=\omega_y=2\pi\times1.26~\mathrm{MHz},
$
and taking \(z\simeq0\) during the expansion,
$\left\langle\frac{\Delta V}{V}\right\rangle
=
2.74\times10^{-6}.
$
This demonstrates that the correction arising from the transverse motion is negligible, thereby validating the one-dimensional interaction model employed throughout this work.

\bibliography{reference.bib}{}

\begin{thebibliography}{40}%
\makeatletter
\providecommand \@ifxundefined [1]{%
 \@ifx{#1\undefined}
}%
\providecommand \@ifnum [1]{%
 \ifnum #1\expandafter \@firstoftwo
 \else \expandafter \@secondoftwo
 \fi
}%
\providecommand \@ifx [1]{%
 \ifx #1\expandafter \@firstoftwo
 \else \expandafter \@secondoftwo
 \fi
}%
\providecommand \natexlab [1]{#1}%
\providecommand \enquote  [1]{``#1''}%
\providecommand \bibnamefont  [1]{#1}%
\providecommand \bibfnamefont [1]{#1}%
\providecommand \citenamefont [1]{#1}%
\providecommand \href@noop [0]{\@secondoftwo}%
\providecommand \href [0]{\begingroup \@sanitize@url \@href}%
\providecommand \@href[1]{\@@startlink{#1}\@@href}%
\providecommand \@@href[1]{\endgroup#1\@@endlink}%
\providecommand \@sanitize@url [0]{\catcode `\\12\catcode `\$12\catcode `\&12\catcode `\#12\catcode `\^12\catcode `\_12\catcode `\%12\relax}%
\providecommand \@@startlink[1]{}%
\providecommand \@@endlink[0]{}%
\providecommand \url  [0]{\begingroup\@sanitize@url \@url }%
\providecommand \@url [1]{\endgroup\@href {#1}{\urlprefix }}%
\providecommand \urlprefix  [0]{URL }%
\providecommand \Eprint [0]{\href }%
\providecommand \doibase [0]{https://doi.org/}%
\providecommand \selectlanguage [0]{\@gobble}%
\providecommand \bibinfo  [0]{\@secondoftwo}%
\providecommand \bibfield  [0]{\@secondoftwo}%
\providecommand \translation [1]{[#1]}%
\providecommand \BibitemOpen [0]{}%
\providecommand \bibitemStop [0]{}%
\providecommand \bibitemNoStop [0]{.\EOS\space}%
\providecommand \EOS [0]{\spacefactor3000\relax}%
\providecommand \BibitemShut  [1]{\csname bibitem#1\endcsname}%
\let\auto@bib@innerbib\@empty
\bibitem [{\citenamefont {Campbell}\ \emph {et~al.}(2017)\citenamefont {Campbell}, \citenamefont {Terhal},\ and\ \citenamefont {Vuillot}}]{campbell2017roads}%
  \BibitemOpen
  \bibfield  {author} {\bibinfo {author} {\bibfnamefont {E.~T.}\ \bibnamefont {Campbell}}, \bibinfo {author} {\bibfnamefont {B.~M.}\ \bibnamefont {Terhal}},\ and\ \bibinfo {author} {\bibfnamefont {C.}~\bibnamefont {Vuillot}},\ }\href {https://doi.org/10.1038/nature23460} {\bibfield  {journal} {\bibinfo  {journal} {Nature}\ }\textbf {\bibinfo {volume} {549}},\ \bibinfo {pages} {172} (\bibinfo {year} {2017})}\BibitemShut {NoStop}%
\bibitem [{\citenamefont {Albert}\ \emph {et~al.}(2018)\citenamefont {Albert}, \citenamefont {Noh}, \citenamefont {Duivenvoorden}, \citenamefont {Young}, \citenamefont {Brierley}, \citenamefont {Reinhold}, \citenamefont {Vuillot}, \citenamefont {Li}, \citenamefont {Shen}, \citenamefont {Girvin}, \citenamefont {Terhal},\ and\ \citenamefont {Jiang}}]{albert2018performance}%
  \BibitemOpen
  \bibfield  {author} {\bibinfo {author} {\bibfnamefont {V.~V.}\ \bibnamefont {Albert}}, \bibinfo {author} {\bibfnamefont {K.}~\bibnamefont {Noh}}, \bibinfo {author} {\bibfnamefont {K.}~\bibnamefont {Duivenvoorden}}, \bibinfo {author} {\bibfnamefont {D.~J.}\ \bibnamefont {Young}}, \bibinfo {author} {\bibfnamefont {R.~T.}\ \bibnamefont {Brierley}}, \bibinfo {author} {\bibfnamefont {P.}~\bibnamefont {Reinhold}}, \bibinfo {author} {\bibfnamefont {C.}~\bibnamefont {Vuillot}}, \bibinfo {author} {\bibfnamefont {L.}~\bibnamefont {Li}}, \bibinfo {author} {\bibfnamefont {C.}~\bibnamefont {Shen}}, \bibinfo {author} {\bibfnamefont {S.~M.}\ \bibnamefont {Girvin}}, \bibinfo {author} {\bibfnamefont {B.~M.}\ \bibnamefont {Terhal}},\ and\ \bibinfo {author} {\bibfnamefont {L.}~\bibnamefont {Jiang}},\ }\href {https://doi.org/10.1103/PhysRevA.97.032346} {\bibfield  {journal} {\bibinfo  {journal} {Phys. Rev. A}\ }\textbf {\bibinfo {volume} {97}},\ \bibinfo {pages} {032346} (\bibinfo {year} {2018})}\BibitemShut {NoStop}%
\bibitem [{\citenamefont {Gottesman}\ \emph {et~al.}(2001)\citenamefont {Gottesman}, \citenamefont {Kitaev},\ and\ \citenamefont {Preskill}}]{gottesman2001encoding}%
  \BibitemOpen
  \bibfield  {author} {\bibinfo {author} {\bibfnamefont {D.}~\bibnamefont {Gottesman}}, \bibinfo {author} {\bibfnamefont {A.}~\bibnamefont {Kitaev}},\ and\ \bibinfo {author} {\bibfnamefont {J.}~\bibnamefont {Preskill}},\ }\href {https://doi.org/10.1103/PhysRevA.64.012310} {\bibfield  {journal} {\bibinfo  {journal} {Phys. Rev. A}\ }\textbf {\bibinfo {volume} {64}},\ \bibinfo {pages} {012310} (\bibinfo {year} {2001})}\BibitemShut {NoStop}%
\bibitem [{\citenamefont {Grimsmo}\ and\ \citenamefont {Puri}(2021)}]{grimsmo2021gkp}%
  \BibitemOpen
  \bibfield  {author} {\bibinfo {author} {\bibfnamefont {A.~L.}\ \bibnamefont {Grimsmo}}\ and\ \bibinfo {author} {\bibfnamefont {S.}~\bibnamefont {Puri}},\ }\href {https://doi.org/10.1103/PRXQuantum.2.020101} {\bibfield  {journal} {\bibinfo  {journal} {PRX Quantum}\ }\textbf {\bibinfo {volume} {2}},\ \bibinfo {pages} {020101} (\bibinfo {year} {2021})}\BibitemShut {NoStop}%
\bibitem [{\citenamefont {Brady}\ \emph {et~al.}(2024)\citenamefont {Brady}, \citenamefont {Eickbusch}, \citenamefont {Singh}, \citenamefont {Wu},\ and\ \citenamefont {Zhuang}}]{brady2024advances}%
  \BibitemOpen
  \bibfield  {author} {\bibinfo {author} {\bibfnamefont {A.~J.}\ \bibnamefont {Brady}}, \bibinfo {author} {\bibfnamefont {A.}~\bibnamefont {Eickbusch}}, \bibinfo {author} {\bibfnamefont {S.}~\bibnamefont {Singh}}, \bibinfo {author} {\bibfnamefont {J.}~\bibnamefont {Wu}},\ and\ \bibinfo {author} {\bibfnamefont {Q.}~\bibnamefont {Zhuang}},\ }\href {https://doi.org/10.1016/j.pquantelec.2023.100496} {\bibfield  {journal} {\bibinfo  {journal} {Prog. Quantum Electron.}\ }\textbf {\bibinfo {volume} {93}},\ \bibinfo {pages} {100496} (\bibinfo {year} {2024})}\BibitemShut {NoStop}%
\bibitem [{\citenamefont {Campagne-Ibarcq}\ \emph {et~al.}(2020)\citenamefont {Campagne-Ibarcq}, \citenamefont {Eickbusch}, \citenamefont {Touzard}, \citenamefont {Zalys-Geller}, \citenamefont {Frattini}, \citenamefont {Sivak}, \citenamefont {Reinhold}, \citenamefont {Puri}, \citenamefont {Shankar}, \citenamefont {Schoelkopf}, \citenamefont {Frunzio}, \citenamefont {Mirrahimi},\ and\ \citenamefont {Devoret}}]{campagne-ibarcq2020qec}%
  \BibitemOpen
  \bibfield  {author} {\bibinfo {author} {\bibfnamefont {P.}~\bibnamefont {Campagne-Ibarcq}}, \bibinfo {author} {\bibfnamefont {A.}~\bibnamefont {Eickbusch}}, \bibinfo {author} {\bibfnamefont {S.}~\bibnamefont {Touzard}}, \bibinfo {author} {\bibfnamefont {E.}~\bibnamefont {Zalys-Geller}}, \bibinfo {author} {\bibfnamefont {N.~E.}\ \bibnamefont {Frattini}}, \bibinfo {author} {\bibfnamefont {V.~V.}\ \bibnamefont {Sivak}}, \bibinfo {author} {\bibfnamefont {P.}~\bibnamefont {Reinhold}}, \bibinfo {author} {\bibfnamefont {S.}~\bibnamefont {Puri}}, \bibinfo {author} {\bibfnamefont {S.}~\bibnamefont {Shankar}}, \bibinfo {author} {\bibfnamefont {R.~J.}\ \bibnamefont {Schoelkopf}}, \bibinfo {author} {\bibfnamefont {L.}~\bibnamefont {Frunzio}}, \bibinfo {author} {\bibfnamefont {M.}~\bibnamefont {Mirrahimi}},\ and\ \bibinfo {author} {\bibfnamefont {M.~H.}\ \bibnamefont {Devoret}},\ }\href {https://doi.org/10.1038/s41586-020-2603-3} {\bibfield  {journal} {\bibinfo  {journal} {Nature}\ }\textbf {\bibinfo {volume} {584}},\
  \bibinfo {pages} {368} (\bibinfo {year} {2020})}\BibitemShut {NoStop}%
\bibitem [{\citenamefont {Eickbusch}\ \emph {et~al.}(2022)\citenamefont {Eickbusch}, \citenamefont {Sivak}, \citenamefont {Ding}, \citenamefont {Elder}, \citenamefont {Jha}, \citenamefont {Venkatraman}, \citenamefont {Royer}, \citenamefont {Girvin}, \citenamefont {Schoelkopf},\ and\ \citenamefont {Devoret}}]{eickbusch2022fast}%
  \BibitemOpen
  \bibfield  {author} {\bibinfo {author} {\bibfnamefont {A.}~\bibnamefont {Eickbusch}}, \bibinfo {author} {\bibfnamefont {V.}~\bibnamefont {Sivak}}, \bibinfo {author} {\bibfnamefont {A.~Z.}\ \bibnamefont {Ding}}, \bibinfo {author} {\bibfnamefont {S.~S.}\ \bibnamefont {Elder}}, \bibinfo {author} {\bibfnamefont {S.~R.}\ \bibnamefont {Jha}}, \bibinfo {author} {\bibfnamefont {J.}~\bibnamefont {Venkatraman}}, \bibinfo {author} {\bibfnamefont {B.}~\bibnamefont {Royer}}, \bibinfo {author} {\bibfnamefont {S.~M.}\ \bibnamefont {Girvin}}, \bibinfo {author} {\bibfnamefont {R.~J.}\ \bibnamefont {Schoelkopf}},\ and\ \bibinfo {author} {\bibfnamefont {M.~H.}\ \bibnamefont {Devoret}},\ }\href {https://doi.org/10.1038/s41567-022-01776-9} {\bibfield  {journal} {\bibinfo  {journal} {Nat. Phys.}\ }\textbf {\bibinfo {volume} {18}},\ \bibinfo {pages} {1464} (\bibinfo {year} {2022})}\BibitemShut {NoStop}%
\bibitem [{\citenamefont {Sivak}\ \emph {et~al.}(2023)\citenamefont {Sivak}, \citenamefont {Eickbusch}, \citenamefont {Royer}, \citenamefont {Singh}, \citenamefont {Tsioutsios}, \citenamefont {Ganjam}, \citenamefont {Miano}, \citenamefont {Brock}, \citenamefont {Ding}, \citenamefont {Frunzio}, \citenamefont {Girvin}, \citenamefont {Schoelkopf},\ and\ \citenamefont {Devoret}}]{sivak2023realtime}%
  \BibitemOpen
  \bibfield  {author} {\bibinfo {author} {\bibfnamefont {V.~V.}\ \bibnamefont {Sivak}}, \bibinfo {author} {\bibfnamefont {A.}~\bibnamefont {Eickbusch}}, \bibinfo {author} {\bibfnamefont {B.}~\bibnamefont {Royer}}, \bibinfo {author} {\bibfnamefont {S.}~\bibnamefont {Singh}}, \bibinfo {author} {\bibfnamefont {I.}~\bibnamefont {Tsioutsios}}, \bibinfo {author} {\bibfnamefont {S.}~\bibnamefont {Ganjam}}, \bibinfo {author} {\bibfnamefont {A.}~\bibnamefont {Miano}}, \bibinfo {author} {\bibfnamefont {B.~L.}\ \bibnamefont {Brock}}, \bibinfo {author} {\bibfnamefont {A.~Z.}\ \bibnamefont {Ding}}, \bibinfo {author} {\bibfnamefont {L.}~\bibnamefont {Frunzio}}, \bibinfo {author} {\bibfnamefont {S.~M.}\ \bibnamefont {Girvin}}, \bibinfo {author} {\bibfnamefont {R.~J.}\ \bibnamefont {Schoelkopf}},\ and\ \bibinfo {author} {\bibfnamefont {M.~H.}\ \bibnamefont {Devoret}},\ }\href {https://doi.org/10.1038/s41586-023-05782-6} {\bibfield  {journal} {\bibinfo  {journal} {Nature}\ }\textbf {\bibinfo {volume} {616}},\ \bibinfo {pages}
  {50} (\bibinfo {year} {2023})}\BibitemShut {NoStop}%
\bibitem [{\citenamefont {Lachance-Quirion}\ \emph {et~al.}(2024)\citenamefont {Lachance-Quirion}, \citenamefont {Lemonde}, \citenamefont {Simoneau}, \citenamefont {St-Jean}, \citenamefont {Lemieux}, \citenamefont {Turcotte}, \citenamefont {Wright}, \citenamefont {Lacroix}, \citenamefont {Fr{\'e}chette-Viens}, \citenamefont {Shillito}, \citenamefont {Hopfmueller}, \citenamefont {Tremblay}, \citenamefont {Frattini}, \citenamefont {Camirand~Lemyre},\ and\ \citenamefont {St-Jean}}]{lachance-quirion2024autonomous}%
  \BibitemOpen
  \bibfield  {author} {\bibinfo {author} {\bibfnamefont {D.}~\bibnamefont {Lachance-Quirion}}, \bibinfo {author} {\bibfnamefont {M.-A.}\ \bibnamefont {Lemonde}}, \bibinfo {author} {\bibfnamefont {J.~O.}\ \bibnamefont {Simoneau}}, \bibinfo {author} {\bibfnamefont {L.}~\bibnamefont {St-Jean}}, \bibinfo {author} {\bibfnamefont {P.}~\bibnamefont {Lemieux}}, \bibinfo {author} {\bibfnamefont {S.}~\bibnamefont {Turcotte}}, \bibinfo {author} {\bibfnamefont {W.}~\bibnamefont {Wright}}, \bibinfo {author} {\bibfnamefont {A.}~\bibnamefont {Lacroix}}, \bibinfo {author} {\bibfnamefont {J.}~\bibnamefont {Fr{\'e}chette-Viens}}, \bibinfo {author} {\bibfnamefont {R.}~\bibnamefont {Shillito}}, \bibinfo {author} {\bibfnamefont {F.}~\bibnamefont {Hopfmueller}}, \bibinfo {author} {\bibfnamefont {M.}~\bibnamefont {Tremblay}}, \bibinfo {author} {\bibfnamefont {N.~E.}\ \bibnamefont {Frattini}}, \bibinfo {author} {\bibfnamefont {J.}~\bibnamefont {Camirand~Lemyre}},\ and\ \bibinfo {author} {\bibfnamefont {P.}~\bibnamefont {St-Jean}},\
  }\href {https://doi.org/10.1103/PhysRevLett.132.150607} {\bibfield  {journal} {\bibinfo  {journal} {Phys. Rev. Lett.}\ }\textbf {\bibinfo {volume} {132}},\ \bibinfo {pages} {150607} (\bibinfo {year} {2024})}\BibitemShut {NoStop}%
\bibitem [{\citenamefont {Fl\"uhmann}\ \emph {et~al.}(2019)\citenamefont {Fl\"uhmann}, \citenamefont {Nguyen}, \citenamefont {Marinelli}, \citenamefont {Negnevitsky}, \citenamefont {Mehta},\ and\ \citenamefont {Home}}]{fluhmann2019encoding}%
  \BibitemOpen
  \bibfield  {author} {\bibinfo {author} {\bibfnamefont {C.}~\bibnamefont {Fl\"uhmann}}, \bibinfo {author} {\bibfnamefont {T.~L.}\ \bibnamefont {Nguyen}}, \bibinfo {author} {\bibfnamefont {M.}~\bibnamefont {Marinelli}}, \bibinfo {author} {\bibfnamefont {V.}~\bibnamefont {Negnevitsky}}, \bibinfo {author} {\bibfnamefont {K.}~\bibnamefont {Mehta}},\ and\ \bibinfo {author} {\bibfnamefont {J.~P.}\ \bibnamefont {Home}},\ }\href {https://doi.org/10.1038/s41586-019-0960-6} {\bibfield  {journal} {\bibinfo  {journal} {Nature}\ }\textbf {\bibinfo {volume} {566}},\ \bibinfo {pages} {513} (\bibinfo {year} {2019})}\BibitemShut {NoStop}%
\bibitem [{\citenamefont {de~Neeve}\ \emph {et~al.}(2022)\citenamefont {de~Neeve}, \citenamefont {Nguyen}, \citenamefont {Behrle},\ and\ \citenamefont {Home}}]{deneeve2022errorcorrection}%
  \BibitemOpen
  \bibfield  {author} {\bibinfo {author} {\bibfnamefont {B.}~\bibnamefont {de~Neeve}}, \bibinfo {author} {\bibfnamefont {T.-L.}\ \bibnamefont {Nguyen}}, \bibinfo {author} {\bibfnamefont {T.}~\bibnamefont {Behrle}},\ and\ \bibinfo {author} {\bibfnamefont {J.~P.}\ \bibnamefont {Home}},\ }\href {https://doi.org/10.1038/s41567-021-01487-7} {\bibfield  {journal} {\bibinfo  {journal} {Nat. Phys.}\ }\textbf {\bibinfo {volume} {18}},\ \bibinfo {pages} {296} (\bibinfo {year} {2022})}\BibitemShut {NoStop}%
\bibitem [{\citenamefont {Matsos}\ \emph {et~al.}(2024)\citenamefont {Matsos}, \citenamefont {Valahu}, \citenamefont {Navickas}, \citenamefont {Rao}, \citenamefont {Millican}, \citenamefont {Kolesnikow}, \citenamefont {Biercuk},\ and\ \citenamefont {Tan}}]{matsos2024robust}%
  \BibitemOpen
  \bibfield  {author} {\bibinfo {author} {\bibfnamefont {V.~G.}\ \bibnamefont {Matsos}}, \bibinfo {author} {\bibfnamefont {C.~H.}\ \bibnamefont {Valahu}}, \bibinfo {author} {\bibfnamefont {T.}~\bibnamefont {Navickas}}, \bibinfo {author} {\bibfnamefont {A.~D.}\ \bibnamefont {Rao}}, \bibinfo {author} {\bibfnamefont {M.~J.}\ \bibnamefont {Millican}}, \bibinfo {author} {\bibfnamefont {X.~C.}\ \bibnamefont {Kolesnikow}}, \bibinfo {author} {\bibfnamefont {M.~J.}\ \bibnamefont {Biercuk}},\ and\ \bibinfo {author} {\bibfnamefont {T.~R.}\ \bibnamefont {Tan}},\ }\href {https://doi.org/10.1103/PhysRevLett.133.050602} {\bibfield  {journal} {\bibinfo  {journal} {Phys. Rev. Lett.}\ }\textbf {\bibinfo {volume} {133}},\ \bibinfo {pages} {050602} (\bibinfo {year} {2024})}\BibitemShut {NoStop}%
\bibitem [{\citenamefont {Matsos}\ \emph {et~al.}(2025)\citenamefont {Matsos}, \citenamefont {Valahu}, \citenamefont {Millican}, \citenamefont {Navickas}, \citenamefont {Kolesnikow}, \citenamefont {Biercuk},\ and\ \citenamefont {Tan}}]{Matsos2025}%
  \BibitemOpen
  \bibfield  {author} {\bibinfo {author} {\bibfnamefont {V.~G.}\ \bibnamefont {Matsos}}, \bibinfo {author} {\bibfnamefont {C.}~\bibnamefont {Valahu}}, \bibinfo {author} {\bibfnamefont {M.~J.}\ \bibnamefont {Millican}}, \bibinfo {author} {\bibfnamefont {T.}~\bibnamefont {Navickas}}, \bibinfo {author} {\bibfnamefont {X.~C.}\ \bibnamefont {Kolesnikow}}, \bibinfo {author} {\bibfnamefont {M.~J.}\ \bibnamefont {Biercuk}},\ and\ \bibinfo {author} {\bibfnamefont {T.~R.}\ \bibnamefont {Tan}},\ }\href {https://doi.org/10.1038/s41567-025-03002-8} {\bibfield  {journal} {\bibinfo  {journal} {Nature Physics}\ }\textbf {\bibinfo {volume} {21}},\ \bibinfo {pages} {1664} (\bibinfo {year} {2025})}\BibitemShut {NoStop}%
\bibitem [{\citenamefont {Rojkov}\ \emph {et~al.}(2024)\citenamefont {Rojkov}, \citenamefont {R{\"o}ggla}, \citenamefont {Wagener}, \citenamefont {Fontbot{\'e}-Schmidt}, \citenamefont {Welte}, \citenamefont {Home},\ and\ \citenamefont {Reiter}}]{PhysRevLett.133.100601}%
  \BibitemOpen
  \bibfield  {author} {\bibinfo {author} {\bibfnamefont {I.}~\bibnamefont {Rojkov}}, \bibinfo {author} {\bibfnamefont {P.~M.}\ \bibnamefont {R{\"o}ggla}}, \bibinfo {author} {\bibfnamefont {M.}~\bibnamefont {Wagener}}, \bibinfo {author} {\bibfnamefont {M.}~\bibnamefont {Fontbot{\'e}-Schmidt}}, \bibinfo {author} {\bibfnamefont {S.}~\bibnamefont {Welte}}, \bibinfo {author} {\bibfnamefont {J.}~\bibnamefont {Home}},\ and\ \bibinfo {author} {\bibfnamefont {F.}~\bibnamefont {Reiter}},\ }\href {https://doi.org/10.1103/PhysRevLett.133.100601} {\bibfield  {journal} {\bibinfo  {journal} {Phys. Rev. Lett.}\ }\textbf {\bibinfo {volume} {133}},\ \bibinfo {pages} {100601} (\bibinfo {year} {2024})}\BibitemShut {NoStop}%
\bibitem [{\citenamefont {Saffman}(2016)}]{Saffman2016}%
  \BibitemOpen
  \bibfield  {author} {\bibinfo {author} {\bibfnamefont {M.}~\bibnamefont {Saffman}},\ }\href {https://doi.org/10.1088/0953-4075/49/20/202001} {\bibfield  {journal} {\bibinfo  {journal} {J. Phys. B: At. Mol. Opt. Phys.}\ }\textbf {\bibinfo {volume} {49}},\ \bibinfo {pages} {202001} (\bibinfo {year} {2016})}\BibitemShut {NoStop}%
\bibitem [{\citenamefont {Henriet}\ \emph {et~al.}(2020)\citenamefont {Henriet}, \citenamefont {Beguin}, \citenamefont {Signoles}, \citenamefont {Lahaye}, \citenamefont {Browaeys}, \citenamefont {Reymond},\ and\ \citenamefont {Jurczak}}]{Henriet2020quantumcomputing}%
  \BibitemOpen
  \bibfield  {author} {\bibinfo {author} {\bibfnamefont {L.}~\bibnamefont {Henriet}}, \bibinfo {author} {\bibfnamefont {L.}~\bibnamefont {Beguin}}, \bibinfo {author} {\bibfnamefont {A.}~\bibnamefont {Signoles}}, \bibinfo {author} {\bibfnamefont {T.}~\bibnamefont {Lahaye}}, \bibinfo {author} {\bibfnamefont {A.}~\bibnamefont {Browaeys}}, \bibinfo {author} {\bibfnamefont {G.-O.}\ \bibnamefont {Reymond}},\ and\ \bibinfo {author} {\bibfnamefont {C.}~\bibnamefont {Jurczak}},\ }\href {https://doi.org/10.22331/q-2020-09-21-327} {\bibfield  {journal} {\bibinfo  {journal} {{Quantum}}\ }\textbf {\bibinfo {volume} {4}},\ \bibinfo {pages} {327} (\bibinfo {year} {2020})}\BibitemShut {NoStop}%
\bibitem [{\citenamefont {Bluvstein}\ \emph {et~al.}(2024)\citenamefont {Bluvstein}, \citenamefont {Evered}, \citenamefont {Geim}, \citenamefont {Li}, \citenamefont {Zhou}, \citenamefont {Manovitz}, \citenamefont {Ebadi}, \citenamefont {Cain}, \citenamefont {Kalinowski}, \citenamefont {Hangleiter} \emph {et~al.}}]{bluvstein2024logical}%
  \BibitemOpen
  \bibfield  {author} {\bibinfo {author} {\bibfnamefont {D.}~\bibnamefont {Bluvstein}}, \bibinfo {author} {\bibfnamefont {S.~J.}\ \bibnamefont {Evered}}, \bibinfo {author} {\bibfnamefont {A.~A.}\ \bibnamefont {Geim}}, \bibinfo {author} {\bibfnamefont {S.~H.}\ \bibnamefont {Li}}, \bibinfo {author} {\bibfnamefont {H.}~\bibnamefont {Zhou}}, \bibinfo {author} {\bibfnamefont {T.}~\bibnamefont {Manovitz}}, \bibinfo {author} {\bibfnamefont {S.}~\bibnamefont {Ebadi}}, \bibinfo {author} {\bibfnamefont {M.}~\bibnamefont {Cain}}, \bibinfo {author} {\bibfnamefont {M.}~\bibnamefont {Kalinowski}}, \bibinfo {author} {\bibfnamefont {D.}~\bibnamefont {Hangleiter}}, \emph {et~al.},\ }\href {https://doi.org/10.1038/s41586-023-06927-3} {\bibfield  {journal} {\bibinfo  {journal} {Nature}\ }\textbf {\bibinfo {volume} {626}},\ \bibinfo {pages} {58} (\bibinfo {year} {2024})}\BibitemShut {NoStop}%
\bibitem [{\citenamefont {Morinaga}\ \emph {et~al.}(1999)\citenamefont {Morinaga}, \citenamefont {Bouchoule}, \citenamefont {Karam},\ and\ \citenamefont {Salomon}}]{morinaga1999manipulation}%
  \BibitemOpen
  \bibfield  {author} {\bibinfo {author} {\bibfnamefont {M.}~\bibnamefont {Morinaga}}, \bibinfo {author} {\bibfnamefont {I.}~\bibnamefont {Bouchoule}}, \bibinfo {author} {\bibfnamefont {J.-C.}\ \bibnamefont {Karam}},\ and\ \bibinfo {author} {\bibfnamefont {C.}~\bibnamefont {Salomon}},\ }\href {https://doi.org/10.1103/PhysRevLett.83.4037} {\bibfield  {journal} {\bibinfo  {journal} {Phys. Rev. Lett.}\ }\textbf {\bibinfo {volume} {83}},\ \bibinfo {pages} {4037} (\bibinfo {year} {1999})}\BibitemShut {NoStop}%
\bibitem [{\citenamefont {Bouchoule}\ \emph {et~al.}(1999)\citenamefont {Bouchoule}, \citenamefont {Perrin}, \citenamefont {Kuhn}, \citenamefont {Morinaga},\ and\ \citenamefont {Salomon}}]{bouchoule1999neutral}%
  \BibitemOpen
  \bibfield  {author} {\bibinfo {author} {\bibfnamefont {I.}~\bibnamefont {Bouchoule}}, \bibinfo {author} {\bibfnamefont {H.}~\bibnamefont {Perrin}}, \bibinfo {author} {\bibfnamefont {A.}~\bibnamefont {Kuhn}}, \bibinfo {author} {\bibfnamefont {M.}~\bibnamefont {Morinaga}},\ and\ \bibinfo {author} {\bibfnamefont {C.}~\bibnamefont {Salomon}},\ }\href {https://doi.org/10.1103/PhysRevA.59.R8} {\bibfield  {journal} {\bibinfo  {journal} {Phys. Rev. A}\ }\textbf {\bibinfo {volume} {59}},\ \bibinfo {pages} {R8} (\bibinfo {year} {1999})}\BibitemShut {NoStop}%
\bibitem [{\citenamefont {Belmechri}\ \emph {et~al.}(2013)\citenamefont {Belmechri}, \citenamefont {F\"orster}, \citenamefont {Alt}, \citenamefont {Widera}, \citenamefont {Meschede},\ and\ \citenamefont {Alberti}}]{belmechri2013microwave}%
  \BibitemOpen
  \bibfield  {author} {\bibinfo {author} {\bibfnamefont {N.}~\bibnamefont {Belmechri}}, \bibinfo {author} {\bibfnamefont {L.}~\bibnamefont {F\"orster}}, \bibinfo {author} {\bibfnamefont {W.}~\bibnamefont {Alt}}, \bibinfo {author} {\bibfnamefont {A.}~\bibnamefont {Widera}}, \bibinfo {author} {\bibfnamefont {D.}~\bibnamefont {Meschede}},\ and\ \bibinfo {author} {\bibfnamefont {A.}~\bibnamefont {Alberti}},\ }\href {https://doi.org/10.1088/0953-4075/46/10/104006} {\bibfield  {journal} {\bibinfo  {journal} {J. Phys. B: At. Mol. Opt. Phys.}\ }\textbf {\bibinfo {volume} {46}},\ \bibinfo {pages} {104006} (\bibinfo {year} {2013})}\BibitemShut {NoStop}%
\bibitem [{\citenamefont {Winkelmann}\ \emph {et~al.}(2022)\citenamefont {Winkelmann}, \citenamefont {Weidner}, \citenamefont {Ramola}, \citenamefont {Alt}, \citenamefont {Meschede},\ and\ \citenamefont {Alberti}}]{winkelmann2022direct}%
  \BibitemOpen
  \bibfield  {author} {\bibinfo {author} {\bibfnamefont {F.-R.}\ \bibnamefont {Winkelmann}}, \bibinfo {author} {\bibfnamefont {C.~A.}\ \bibnamefont {Weidner}}, \bibinfo {author} {\bibfnamefont {G.}~\bibnamefont {Ramola}}, \bibinfo {author} {\bibfnamefont {W.}~\bibnamefont {Alt}}, \bibinfo {author} {\bibfnamefont {D.}~\bibnamefont {Meschede}},\ and\ \bibinfo {author} {\bibfnamefont {A.}~\bibnamefont {Alberti}},\ }\href {https://doi.org/10.1088/1361-6455/ac8bb8} {\bibfield  {journal} {\bibinfo  {journal} {J. Phys. B: At. Mol. Opt. Phys.}\ }\textbf {\bibinfo {volume} {55}},\ \bibinfo {pages} {194004} (\bibinfo {year} {2022})}\BibitemShut {NoStop}%
\bibitem [{\citenamefont {Kendell}\ \emph {et~al.}(2024)\citenamefont {Kendell}, \citenamefont {Ferranti},\ and\ \citenamefont {Weidner}}]{kendell2024deterministic}%
  \BibitemOpen
  \bibfield  {author} {\bibinfo {author} {\bibfnamefont {H.~C.}\ \bibnamefont {Kendell}}, \bibinfo {author} {\bibfnamefont {G.}~\bibnamefont {Ferranti}},\ and\ \bibinfo {author} {\bibfnamefont {C.~A.}\ \bibnamefont {Weidner}},\ }\href {https://doi.org/10.1063/5.0197119} {\bibfield  {journal} {\bibinfo  {journal} {APL Quantum}\ }\textbf {\bibinfo {volume} {1}},\ \bibinfo {pages} {026109} (\bibinfo {year} {2024})}\BibitemShut {NoStop}%
\bibitem [{\citenamefont {Grochowski}\ \emph {et~al.}(2025)\citenamefont {Grochowski}, \citenamefont {Pichler}, \citenamefont {Regal},\ and\ \citenamefont {Romero-Isart}}]{Grochowski2025quantumcontrolof}%
  \BibitemOpen
  \bibfield  {author} {\bibinfo {author} {\bibfnamefont {P.~T.}\ \bibnamefont {Grochowski}}, \bibinfo {author} {\bibfnamefont {H.}~\bibnamefont {Pichler}}, \bibinfo {author} {\bibfnamefont {C.~A.}\ \bibnamefont {Regal}},\ and\ \bibinfo {author} {\bibfnamefont {O.}~\bibnamefont {Romero-Isart}},\ }\href {https://doi.org/10.22331/q-2025-08-08-1824} {\bibfield  {journal} {\bibinfo  {journal} {{Quantum}}\ }\textbf {\bibinfo {volume} {9}},\ \bibinfo {pages} {1824} (\bibinfo {year} {2025})}\BibitemShut {NoStop}%
\bibitem [{\citenamefont {Brown}\ \emph {et~al.}(2023)\citenamefont {Brown}, \citenamefont {Muleady}, \citenamefont {Dworschack}, \citenamefont {Lewis-Swan}, \citenamefont {Rey}, \citenamefont {Romero-Isart},\ and\ \citenamefont {Regal}}]{brown2023time}%
  \BibitemOpen
  \bibfield  {author} {\bibinfo {author} {\bibfnamefont {M.}~\bibnamefont {Brown}}, \bibinfo {author} {\bibfnamefont {S.}~\bibnamefont {Muleady}}, \bibinfo {author} {\bibfnamefont {W.}~\bibnamefont {Dworschack}}, \bibinfo {author} {\bibfnamefont {R.}~\bibnamefont {Lewis-Swan}}, \bibinfo {author} {\bibfnamefont {A.}~\bibnamefont {Rey}}, \bibinfo {author} {\bibfnamefont {O.}~\bibnamefont {Romero-Isart}},\ and\ \bibinfo {author} {\bibfnamefont {C.}~\bibnamefont {Regal}},\ }\href {https://ui.adsabs.harvard.edu/link_gateway/2023NatPh..19..569B/doi:10.1038/s41567-022-01890-8} {\bibfield  {journal} {\bibinfo  {journal} {Nature Physics}\ }\textbf {\bibinfo {volume} {19}},\ \bibinfo {pages} {569} (\bibinfo {year} {2023})}\BibitemShut {NoStop}%
\bibitem [{\citenamefont {Bohnmann}\ \emph {et~al.}(2025)\citenamefont {Bohnmann}, \citenamefont {Locher}, \citenamefont {Zeiher},\ and\ \citenamefont {M{\"u}ller}}]{bohnmann2025bosonic}%
  \BibitemOpen
  \bibfield  {author} {\bibinfo {author} {\bibfnamefont {L.~H.}\ \bibnamefont {Bohnmann}}, \bibinfo {author} {\bibfnamefont {D.~F.}\ \bibnamefont {Locher}}, \bibinfo {author} {\bibfnamefont {J.}~\bibnamefont {Zeiher}},\ and\ \bibinfo {author} {\bibfnamefont {M.}~\bibnamefont {M{\"u}ller}},\ }\href {https://doi.org/10.1103/PhysRevA.111.022432} {\bibfield  {journal} {\bibinfo  {journal} {Phys. Rev. A}\ }\textbf {\bibinfo {volume} {111}},\ \bibinfo {pages} {022432} (\bibinfo {year} {2025})}\BibitemShut {NoStop}%
\bibitem [{\citenamefont {Noh}\ \emph {et~al.}(2022)\citenamefont {Noh}, \citenamefont {Chamberland},\ and\ \citenamefont {Brand{\~a}o}}]{noh2022surfacegkp}%
  \BibitemOpen
  \bibfield  {author} {\bibinfo {author} {\bibfnamefont {K.}~\bibnamefont {Noh}}, \bibinfo {author} {\bibfnamefont {C.}~\bibnamefont {Chamberland}},\ and\ \bibinfo {author} {\bibfnamefont {F.~G.}\ \bibnamefont {Brand{\~a}o}},\ }\href {https://doi.org/10.1103/PRXQuantum.3.010315} {\bibfield  {journal} {\bibinfo  {journal} {PRX Quantum}\ }\textbf {\bibinfo {volume} {3}},\ \bibinfo {pages} {010315} (\bibinfo {year} {2022})}\BibitemShut {NoStop}%
\bibitem [{\citenamefont {Royer}\ \emph {et~al.}(2020)\citenamefont {Royer}, \citenamefont {Singh},\ and\ \citenamefont {Girvin}}]{royer2020stabilization}%
  \BibitemOpen
  \bibfield  {author} {\bibinfo {author} {\bibfnamefont {B.}~\bibnamefont {Royer}}, \bibinfo {author} {\bibfnamefont {S.}~\bibnamefont {Singh}},\ and\ \bibinfo {author} {\bibfnamefont {S.~M.}\ \bibnamefont {Girvin}},\ }\href {https://doi.org/10.1103/PhysRevLett.125.260509} {\bibfield  {journal} {\bibinfo  {journal} {Phys. Rev. Lett.}\ }\textbf {\bibinfo {volume} {125}},\ \bibinfo {pages} {260509} (\bibinfo {year} {2020})}\BibitemShut {NoStop}%
\bibitem [{\citenamefont {Grimm}\ \emph {et~al.}(2000)\citenamefont {Grimm}, \citenamefont {Weidem{\"u}ller},\ and\ \citenamefont {Ovchinnikov}}]{grimm2000optical}%
  \BibitemOpen
  \bibfield  {author} {\bibinfo {author} {\bibfnamefont {R.}~\bibnamefont {Grimm}}, \bibinfo {author} {\bibfnamefont {M.}~\bibnamefont {Weidem{\"u}ller}},\ and\ \bibinfo {author} {\bibfnamefont {Y.~B.}\ \bibnamefont {Ovchinnikov}},\ }\href {https://doi.org/10.1016/S1049-250X(08)60186-X} {\bibfield  {journal} {\bibinfo  {journal} {Adv. At. Mol. Opt. Phys.}\ }\textbf {\bibinfo {volume} {42}},\ \bibinfo {pages} {95} (\bibinfo {year} {2000})}\BibitemShut {NoStop}%
\bibitem [{\citenamefont {Zhang}\ \emph {et~al.}(2011)\citenamefont {Zhang}, \citenamefont {Robicheaux},\ and\ \citenamefont {Saffman}}]{zhang2011magic}%
  \BibitemOpen
  \bibfield  {author} {\bibinfo {author} {\bibfnamefont {S.}~\bibnamefont {Zhang}}, \bibinfo {author} {\bibfnamefont {F.}~\bibnamefont {Robicheaux}},\ and\ \bibinfo {author} {\bibfnamefont {M.}~\bibnamefont {Saffman}},\ }\href {https://doi.org/10.1103/PhysRevA.84.043408} {\bibfield  {journal} {\bibinfo  {journal} {Phys. Rev. A}\ }\textbf {\bibinfo {volume} {84}},\ \bibinfo {pages} {043408} (\bibinfo {year} {2011})}\BibitemShut {NoStop}%
\bibitem [{\citenamefont {Topcu}\ and\ \citenamefont {Derevianko}(2014)}]{topcu2014divalent}%
  \BibitemOpen
  \bibfield  {author} {\bibinfo {author} {\bibfnamefont {T.}~\bibnamefont {Topcu}}\ and\ \bibinfo {author} {\bibfnamefont {A.}~\bibnamefont {Derevianko}},\ }\href {https://doi.org/10.1103/PhysRevA.89.023411} {\bibfield  {journal} {\bibinfo  {journal} {Phys. Rev. A}\ }\textbf {\bibinfo {volume} {89}},\ \bibinfo {pages} {023411} (\bibinfo {year} {2014})}\BibitemShut {NoStop}%
\bibitem [{\citenamefont {Pucher}\ \emph {et~al.}(2024)\citenamefont {Pucher}, \citenamefont {Kl{\"u}sener}, \citenamefont {Spriestersbach}, \citenamefont {Geiger}, \citenamefont {Schindewolf}, \citenamefont {Bloch},\ and\ \citenamefont {Blatt}}]{PhysRevLett.132.150605}%
  \BibitemOpen
  \bibfield  {author} {\bibinfo {author} {\bibfnamefont {S.}~\bibnamefont {Pucher}}, \bibinfo {author} {\bibfnamefont {V.}~\bibnamefont {Kl{\"u}sener}}, \bibinfo {author} {\bibfnamefont {F.}~\bibnamefont {Spriestersbach}}, \bibinfo {author} {\bibfnamefont {J.}~\bibnamefont {Geiger}}, \bibinfo {author} {\bibfnamefont {A.}~\bibnamefont {Schindewolf}}, \bibinfo {author} {\bibfnamefont {I.}~\bibnamefont {Bloch}},\ and\ \bibinfo {author} {\bibfnamefont {S.}~\bibnamefont {Blatt}},\ }\href {https://doi.org/10.1103/PhysRevLett.132.150605} {\bibfield  {journal} {\bibinfo  {journal} {Phys. Rev. Lett.}\ }\textbf {\bibinfo {volume} {132}},\ \bibinfo {pages} {150605} (\bibinfo {year} {2024})}\BibitemShut {NoStop}%
\bibitem [{\citenamefont {Unnikrishnan}\ \emph {et~al.}(2024)\citenamefont {Unnikrishnan}, \citenamefont {Ilzh\"ofer}, \citenamefont {Scholz}, \citenamefont {H\"olzl}, \citenamefont {G\"otzelmann}, \citenamefont {Gupta}, \citenamefont {Zhao}, \citenamefont {Krauter}, \citenamefont {Weber}, \citenamefont {Makki}, \citenamefont {B\"uchler}, \citenamefont {Pfau},\ and\ \citenamefont {Meinert}}]{unnikrishnan2024coherent}%
  \BibitemOpen
  \bibfield  {author} {\bibinfo {author} {\bibfnamefont {G.}~\bibnamefont {Unnikrishnan}}, \bibinfo {author} {\bibfnamefont {P.}~\bibnamefont {Ilzh\"ofer}}, \bibinfo {author} {\bibfnamefont {A.}~\bibnamefont {Scholz}}, \bibinfo {author} {\bibfnamefont {C.}~\bibnamefont {H\"olzl}}, \bibinfo {author} {\bibfnamefont {A.}~\bibnamefont {G\"otzelmann}}, \bibinfo {author} {\bibfnamefont {R.~K.}\ \bibnamefont {Gupta}}, \bibinfo {author} {\bibfnamefont {J.}~\bibnamefont {Zhao}}, \bibinfo {author} {\bibfnamefont {J.}~\bibnamefont {Krauter}}, \bibinfo {author} {\bibfnamefont {S.}~\bibnamefont {Weber}}, \bibinfo {author} {\bibfnamefont {N.}~\bibnamefont {Makki}}, \bibinfo {author} {\bibfnamefont {H.~P.}\ \bibnamefont {B\"uchler}}, \bibinfo {author} {\bibfnamefont {T.}~\bibnamefont {Pfau}},\ and\ \bibinfo {author} {\bibfnamefont {F.}~\bibnamefont {Meinert}},\ }\href {https://doi.org/10.1103/PhysRevLett.132.150606} {\bibfield  {journal} {\bibinfo  {journal} {Phys. Rev. Lett.}\ }\textbf {\bibinfo {volume} {132}},\ \bibinfo
  {pages} {150606} (\bibinfo {year} {2024})}\BibitemShut {NoStop}%
\bibitem [{\citenamefont {Saffman}\ \emph {et~al.}(2010)\citenamefont {Saffman}, \citenamefont {Walker},\ and\ \citenamefont {M{\o}lmer}}]{saffman2010rydberg}%
  \BibitemOpen
  \bibfield  {author} {\bibinfo {author} {\bibfnamefont {M.}~\bibnamefont {Saffman}}, \bibinfo {author} {\bibfnamefont {T.~G.}\ \bibnamefont {Walker}},\ and\ \bibinfo {author} {\bibfnamefont {K.}~\bibnamefont {M{\o}lmer}},\ }\href {https://doi.org/10.1103/RevModPhys.82.2313} {\bibfield  {journal} {\bibinfo  {journal} {Rev. Mod. Phys.}\ }\textbf {\bibinfo {volume} {82}},\ \bibinfo {pages} {2313} (\bibinfo {year} {2010})}\BibitemShut {NoStop}%
\bibitem [{\citenamefont {Pagano}\ \emph {et~al.}(2022)\citenamefont {Pagano}, \citenamefont {Weber}, \citenamefont {Jaschke}, \citenamefont {Pfau}, \citenamefont {Meinert}, \citenamefont {Montangero},\ and\ \citenamefont {B{\"u}chler}}]{PhysRevResearch.4.033019}%
  \BibitemOpen
  \bibfield  {author} {\bibinfo {author} {\bibfnamefont {A.}~\bibnamefont {Pagano}}, \bibinfo {author} {\bibfnamefont {S.}~\bibnamefont {Weber}}, \bibinfo {author} {\bibfnamefont {D.}~\bibnamefont {Jaschke}}, \bibinfo {author} {\bibfnamefont {T.}~\bibnamefont {Pfau}}, \bibinfo {author} {\bibfnamefont {F.}~\bibnamefont {Meinert}}, \bibinfo {author} {\bibfnamefont {S.}~\bibnamefont {Montangero}},\ and\ \bibinfo {author} {\bibfnamefont {H.~P.}\ \bibnamefont {B{\"u}chler}},\ }\href {https://doi.org/10.1103/PhysRevResearch.4.033019} {\bibfield  {journal} {\bibinfo  {journal} {Phys. Rev. Res.}\ }\textbf {\bibinfo {volume} {4}},\ \bibinfo {pages} {033019} (\bibinfo {year} {2022})}\BibitemShut {NoStop}%
\bibitem [{\citenamefont {Meinert}\ \emph {et~al.}(2023)\citenamefont {Meinert}, \citenamefont {Pfau},\ and\ \citenamefont {H{\"o}lzl}}]{meinert2023state}%
  \BibitemOpen
  \bibfield  {author} {\bibinfo {author} {\bibfnamefont {F.}~\bibnamefont {Meinert}}, \bibinfo {author} {\bibfnamefont {T.}~\bibnamefont {Pfau}},\ and\ \bibinfo {author} {\bibfnamefont {C.}~\bibnamefont {H{\"o}lzl}},\ }\href@noop {} {\bibinfo {title} {Quantum computing device, use, and method}},\ \bibinfo {howpublished} {European Patent Application EP4264505A1} (\bibinfo {year} {2023})\BibitemShut {NoStop}%
\bibitem [{\citenamefont {Tao}\ \emph {et~al.}(2024)\citenamefont {Tao}, \citenamefont {Ammenwerth}, \citenamefont {Gyger}, \citenamefont {Bloch},\ and\ \citenamefont {Zeiher}}]{tao2024prl}%
  \BibitemOpen
  \bibfield  {author} {\bibinfo {author} {\bibfnamefont {R.}~\bibnamefont {Tao}}, \bibinfo {author} {\bibfnamefont {M.}~\bibnamefont {Ammenwerth}}, \bibinfo {author} {\bibfnamefont {F.}~\bibnamefont {Gyger}}, \bibinfo {author} {\bibfnamefont {I.}~\bibnamefont {Bloch}},\ and\ \bibinfo {author} {\bibfnamefont {J.}~\bibnamefont {Zeiher}},\ }\href {https://doi.org/10.1103/PhysRevLett.133.013401} {\bibfield  {journal} {\bibinfo  {journal} {Phys. Rev. Lett.}\ }\textbf {\bibinfo {volume} {133}},\ \bibinfo {pages} {013401} (\bibinfo {year} {2024})}\BibitemShut {NoStop}%
\bibitem [{\citenamefont {Robertson}\ \emph {et~al.}(2021)\citenamefont {Robertson}, \citenamefont {{\v{S}}ibali{\'c}}, \citenamefont {Potvliege},\ and\ \citenamefont {Jones}}]{Robertson2021ARC}%
  \BibitemOpen
  \bibfield  {author} {\bibinfo {author} {\bibfnamefont {E.~J.}\ \bibnamefont {Robertson}}, \bibinfo {author} {\bibfnamefont {N.}~\bibnamefont {{\v{S}}ibali{\'c}}}, \bibinfo {author} {\bibfnamefont {R.~M.}\ \bibnamefont {Potvliege}},\ and\ \bibinfo {author} {\bibfnamefont {M.~P.~A.}\ \bibnamefont {Jones}},\ }\href {https://doi.org/10.1016/j.cpc.2020.107814} {\bibfield  {journal} {\bibinfo  {journal} {Computer Physics Communications}\ }\textbf {\bibinfo {volume} {261}},\ \bibinfo {pages} {107814} (\bibinfo {year} {2021})}\BibitemShut {NoStop}%
\bibitem [{\citenamefont {Tsai}\ \emph {et~al.}(2025)\citenamefont {Tsai}, \citenamefont {Sun}, \citenamefont {Shaw}, \citenamefont {Finkelstein},\ and\ \citenamefont {Endres}}]{PRXQuantum.6.010331}%
  \BibitemOpen
  \bibfield  {author} {\bibinfo {author} {\bibfnamefont {R.~B.-S.}\ \bibnamefont {Tsai}}, \bibinfo {author} {\bibfnamefont {X.}~\bibnamefont {Sun}}, \bibinfo {author} {\bibfnamefont {A.~L.}\ \bibnamefont {Shaw}}, \bibinfo {author} {\bibfnamefont {R.}~\bibnamefont {Finkelstein}},\ and\ \bibinfo {author} {\bibfnamefont {M.}~\bibnamefont {Endres}},\ }\href {https://doi.org/10.1103/PRXQuantum.6.010331} {\bibfield  {journal} {\bibinfo  {journal} {PRX Quantum}\ }\textbf {\bibinfo {volume} {6}},\ \bibinfo {pages} {010331} (\bibinfo {year} {2025})}\BibitemShut {NoStop}%
\bibitem [{\citenamefont {Henderson}\ \emph {et~al.}(2009)\citenamefont {Henderson}, \citenamefont {Ryu}, \citenamefont {MacCormick},\ and\ \citenamefont {Boshier}}]{Henderson_2009}%
  \BibitemOpen
  \bibfield  {author} {\bibinfo {author} {\bibfnamefont {K.}~\bibnamefont {Henderson}}, \bibinfo {author} {\bibfnamefont {C.}~\bibnamefont {Ryu}}, \bibinfo {author} {\bibfnamefont {C.}~\bibnamefont {MacCormick}},\ and\ \bibinfo {author} {\bibfnamefont {M.~G.}\ \bibnamefont {Boshier}},\ }\href {https://doi.org/10.1088/1367-2630/11/4/043030} {\bibfield  {journal} {\bibinfo  {journal} {New Journal of Physics}\ }\textbf {\bibinfo {volume} {11}},\ \bibinfo {pages} {043030} (\bibinfo {year} {2009})}\BibitemShut {NoStop}%
\bibitem [{\citenamefont {Pampel}\ \emph {et~al.}(2026)\citenamefont {Pampel}, \citenamefont {Lubin}, \citenamefont {Hewatt}, \citenamefont {McCabe}, \citenamefont {Hwang}, \citenamefont {Muleady}, \citenamefont {Xu}, \citenamefont {Rey},\ and\ \citenamefont {Regal}}]{Pampel_2026}%
  \BibitemOpen
  \bibfield  {author} {\bibinfo {author} {\bibfnamefont {S.~K.}\ \bibnamefont {Pampel}}, \bibinfo {author} {\bibfnamefont {G.}~\bibnamefont {Lubin}}, \bibinfo {author} {\bibfnamefont {D.~P.}\ \bibnamefont {Hewatt}}, \bibinfo {author} {\bibfnamefont {C.}~\bibnamefont {McCabe}}, \bibinfo {author} {\bibfnamefont {J.}~\bibnamefont {Hwang}}, \bibinfo {author} {\bibfnamefont {S.~R.}\ \bibnamefont {Muleady}}, \bibinfo {author} {\bibfnamefont {T.}~\bibnamefont {Xu}}, \bibinfo {author} {\bibfnamefont {A.~M.}\ \bibnamefont {Rey}},\ and\ \bibinfo {author} {\bibfnamefont {C.~A.}\ \bibnamefont {Regal}},\ }\bibfield  {journal} {\bibinfo  {journal} {arXiv preprint arXiv:2607.18579}\ }\href {https://doi.org/10.48550/arXiv.2607.18579} {10.48550/arXiv.2607.18579} (\bibinfo {year} {2026})\BibitemShut {NoStop}%
\end{thebibliography}%
\bibliographystyle{apsrev4-2}
\end{document}